\documentclass[10pt]{article}
 
\usepackage{amsmath}
\usepackage{amssymb}
\usepackage{graphicx}
\usepackage{cite}
\usepackage{color}
\usepackage{url}
\usepackage{setspace}

\usepackage[labelfont=bf,labelsep=period,justification=raggedright]{caption}
\makeatletter
\renewcommand{\@biblabel}[1]{\quad#1.}
\makeatother

\date{}
\begin{document}

\begin{flushleft}
{\Large
\textbf{Cascading failures in spatially-embedded random networks}
}
\\
Andrea Asztalos$^{1,2,3}$\footnote{Current address: NCBI, NLM, NIH Bethesda, Maryland, USA},
Sameet Sreenivasan$^{1,2,3}$\footnote{Corresponding author: sreens@rpi.edu},
Boleslaw K. Szymanski$^{1,2}$
Gyorgy Korniss$^{1,3}$
\\
\bf{1} Social and Cognitive Networks Academic Research Center, Rensselaer Polytechnic Institute, Troy, New York, USA
\\
\bf{2} Department of Computer Science, Rensselaer Polytechnic Institute, Troy, New York, USA
\\
\bf{3} Department of Physics, Applied Physics and Astronomy, Rensselaer Polytechnic Institute, Troy, New York, USA
\\
\end{flushleft}

\section*{Abstract}
Cascading failures constitute an important vulnerability of
interconnected systems. Here we focus on the study of such failures
on networks in which the connectivity of nodes is constrained by
geographical distance. Specifically, we use random geometric graphs
as representative examples of such spatial networks, and study the
properties of cascading failures on them in the presence of
distributed flow. The key finding of this study is that the process
of cascading failures is non-self-averaging on spatial networks, and
thus, aggregate inferences made from analyzing an ensemble of such
networks lead to incorrect conclusions when applied to a single
network, no matter how large the network is. We demonstrate that
this lack of self-averaging disappears with the introduction of a
small fraction of long-range links into the network. We simulate the
well studied preemptive node removal strategy for cascade mitigation
and show that it is largely ineffective in the case of spatial
networks. We introduce an altruistic strategy designed to limit the
loss of network nodes in the event of a cascade triggering failure
and show that it performs better than the preemptive strategy.
Finally, we consider a real-world spatial network viz. a European
power transmission network and validate that our findings from the
study of random geometric graphs are also borne out by simulations
of cascading failures on the empirical network.

\section*{Introduction}

Cascading failures represent a particular vulnerability of networked
systems, and their effects have been experienced and documented in
several domains such as infrastructure networks \cite{blackout},
financial systems \cite{Sachs2009}, and biological systems
\cite{Ischemic2011}. An important feature of real-world networks
that has not been incorporated into most studies on cascading
failures, with some notable exceptions \cite{Zussman2011, Verma2013,
Havlin2012}, is the fact that they are subject to spatial
constraints. In other words, in most real-world networks, which node a
given node connects to, or interacts with, is largely determined by
the geographic distance between the two nodes. This rather severe constraint has important consequences on
the network's behavior, and gives rise to significant differences in
the scaling behavior of quantities of interest when compared to
spatially unconstrained networks \cite{Barthelemy2011}.

In the context of cascading failures and strategies for their
mitigation, studying the effect of spatial constraints is crucial to
providing fundamental insights that are practically applicable. A
specific context within which studies of cascading failures have
proliferated is that of electrical power transmission systems
\cite{Kinney05,Dobson07,Zussman2011,Verma2013,Chertkov2013,Scala12,
Majeed2013}. However, understanding such failures in the more
general context of flow bearing networks is just as important,
especially when confronted with the imminent rise of technologies
like the {\it Internet of Things} \cite{Pereira2013}, which
essentially consists of everyday physical objects equipped
with sensors to communicate with users or other objects within
their range.

Motivated by these reasons, we study a model of load-based cascading
failures on networks on a particular class of spatially constrained
networks - the Random Geometric Graph (RGG) \cite{Penrose03,Dall02}
- carrying distributed flows and compare its behavior to that of
unconstrained network classes. Closely related earlier and recent works,
employing resistor networks, investigated transport efficiency, flow
optimization, and vulnerability in complex networks
\cite{Lopez_PRL2005,KornissSWPhysLettA06,KornissweightsPRE,Asztalos12,KornissBookCh},
and the emergence of traffic gridlocks and congestion in road
networks \cite{Mendes_PhysicaA2012,Colak_NJP2013}.

To validate the insights obtained from these spatially-embedded
model networks (RGGs), we also study the same load-based cascading
failure process on a real-world network with spatial constraints -
the European power transmission network maintained and operated by
the Union for the Co-ordination of Transmission of Electricity (UCTE).
We find several revealing features that arise from the presence of
spatial constraints, the most noticeable being a lack of
self-averaging on such networks. This is in stark contrast to the
results for unconstrained random networks, and thus points to the
potential pitfalls of ignoring spatial constraints when studying
cascade mitigation strategies.

\section*{Methods} \label{methods}

Here, we briefly describe the distributed flow model and cascade
model that we utilize in this study. For clarity, we note that we
use the term `node' and `vertex' interchangeably in the rest of the
paper.

\subsection*{1.~Distributed flow}
We assume the flow on the network to be both directed and
distributed. Specifically, each unit of flow is associated with a
source and a sink, and takes advantage of all possible paths between
the source and the sink. We adopt a simple model of such flow, by
modeling the network as a random resistor network with unit
conductances along the links \cite{Asztalos12,KornissBookCh}. As
each node and edge plays a role in forwarding the current from the
source to the target node, each of them experiences a load.  For one
source-target pair and for unit current flowing between them, the
load on an arbitrary edge $e \equiv (i,j)$ is the current along that
edge: $\ell_{ij}=I_{ij}^{(st)}$; analogously, the load on an
arbitrary node $i$ is the net current flowing through that node:
$\ell_{i}=I_{i}^{(st)}$. These two loads are related by the
expression
\begin{equation}
I_i^{(st)} =\frac{1}{2} \sum_{j } |I_{ij}^{(st)} |.
\label{eq:1}
\end{equation}
For all our studies presented here, we assume that unit current
flows between $N$ source/target pairs simultaneously. Specifically,
we assume that all nodes are simultaneously sources and unit current
flows into the network at each source. For each source node, a
target is chosen randomly and uniformly from the remaining $N-1$
nodes. Consequently, the load is defined as the superposition of all
currents flowing through an arbitrary edge/node, which is identical
to the edge/node current-flow betweenness
\cite{Newman_betw,BrandesCflow,KornissBookCh}:
\begin{equation}
\ell_{ij}=\frac{1}{N-1}\sum_{s,t=1}^N |I_{ij}^{(st)}|, \;\; \ell_i =\frac{1}{N-1}\sum_{s,t=1}^N  |I_i^{(st)}|.
\label{eq:2}
\end{equation}
Currents $I_{ij}^{(st)}$ along the edges due to {\it one} source/target pair are obtained by writing down Kirchhoff's law for each node $i$ in the network and solving the system of linear equations:
\begin{equation}
\sum_{j=1}^{N} A_{ij} (V_{i} - V_{j}) = I(\delta_{is} -\delta_{it}), \;\; \forall i=1,\dots,N,
\label{eq:3}
\end{equation}
where we assumed that $I$ units of current flow into the network at a source $s$ and leave at a target $t$, and where $A_{ij}$ denotes the adjacency matrix of the network. Rewritten in terms of the weighted network Laplacian $\mathcal{L}_{ij}=\delta_{ij} k_i - A_{ij}$, where $k_i=\sum_{j}A_{ij}$ denotes the degree of node $i$, the system (\ref{eq:3}) transforms into the matrix equation $\mathcal{L} V=\mathcal{I}$, where $V$ is the unknown column voltage vector, while $\mathcal{I}_i$ is the net current flowing into the network at node $i$, which is zero for all nodes with the exception of the source and target nodes. As the network Laplacian $\mathcal{L}$ is singular, we use spectral decomposition \cite{KornissweightsPRE,Slepc_Hernandez} to find the pseudo-inverse Laplacian $G=\mathcal{L}^{-1}$, defined in the space orthogonal to the zero mode. For example, by choosing the reference potential to be the mean voltage \cite{KornissSWPhysLettA06}, $\hat{V}_i = V_i -\langle V \rangle$, where $\langle V \rangle = (1/N)\sum_{j=1}^N V_j$,  one obtains:
\begin{equation}
\hat{V_i} =(G\mathcal{I})_i =\sum_{j=1}^N G_{ij} I(\delta_{js} -\delta_{jt})= I(G_{is} -G_{it}),
\label{eq:voltage}
\end{equation}
for each node $i$. Thus, for $I$ units of current and for a given source/target pair, the current flowing through edge $(i,j)$ is:
\begin{equation}
I_{ij}^{(st)} = A_{ij}  (V_i -V_j) = A_{ij}I (G_{is}-G_{it}-G_{js}+G_{jt}),
\label{eq:current}
\end{equation}
This relation shows that current along an arbitrary edge is uniquely
determined by the network topology. Therefore, this is a fully
deterministic model of flow and the only source of randomness in the
problem arises in the specific instantiation of the network from its
parent ensemble.

Electrical flows when applied to explicitly modeling the power grid
have commonly used a DC power flow model \cite{Zimmerman,
Zussman2011, Scala12, Majeed2013, Verma2013} wherein links also
possess a reactance in addition to resistance. However, as pointed
out in \cite{Zussman2011}, the equations  for this DC model of power
flow bear a close resemblance to that of a simple electrical circuit
with the current playing the analogous role of power. Further, Scala
et al. \cite{Scala12} have demonstrated that inferences made using a
DC power flow model, can still be useful despite neglecting the true
AC nature of the power transmission network \cite{Helbing08}. We
emphasize that although the empirical network on which we validate our
results is an electrical grid, our studies are aimed at
understanding fundamental aspects of cascades on spatial networks
carrying distributed flow, and not towards designing strategies
specifically tailored for electrical power transmission systems.

\subsection*{2.~Cascade model}
We model a cascading failure on a network carrying distributed flow
following the seminal model of Motter and Lai \cite{MotterLai02}. We
assume that each node is equipped with a load handling capacity that
is proportional to the steady-state load on it when the network is
intact. Specifically, the capacity of a node $i$ is $C_i =
(1+\alpha)\ell_i^0$ where $\alpha$ plays the role of a tolerance
parameter, and $\ell_i^0$ is the load on the node for the intact
network. If a node on the network fails, i.e. is absent or removed
from the network, then the flow undergoes a redistribution, and
consequently, so do the loads on the surviving nodes. If the new
load on any surviving node exceeds its capacity, i.e. if $\ell_i >
C_i$, then that node also fails which leads to a further
redistribution and possibly further failures. This process
constitutes the model of a cascading failure that we utilize here.

\subsection*{3. Network models}

We briefly outline the network models used in this paper and the methods employed for generating associated ensembles.
\subsubsection*{Random Geometric Graphs}
A Random Geometric Graph (RGG) of size $N$ in 2D is constructed by
placing $N$ nodes randomly in the unit square with open boundary
conditions, and connecting any pair of nodes if the Euclidean
distance between them is less than a distance $R$, the {\it
connection radius} \cite{Penrose03,Dall02}. The average degree of
the graph $\langle k \rangle$ can be controlled by varying $R$ since
$\langle k \rangle = \pi R^2 N$.

\subsubsection*{Erd\H{o}s-R\'enyi graphs}
An Erd\H{o}s-R\'enyi (ER) graph \cite{Bollobas} of size $N$ is
constructed by connecting every pair of nodes with probability $p$.
The average degree of the network can be controlled through $p$
since $\langle k \rangle = p (N-1)$.

\subsubsection*{Scale-Free networks}
Scale-free (SF) networks  \cite{BarabasiAlbert99} of size $N$ and
degree-exponent $\gamma$ are constructed by first generating a
degree sequence by sampling the prescribed power-law distribution
$P(k) \sim k^{-\gamma}$ that yields a desired average-degree
$\langle k \rangle$. The network is then constructed using this
degree sequence following the Configuration Model \cite{Molloy95}.

\subsubsection*{Rewired Random Geometric Graphs}
To better understand the role of spatial constraints in the observed
characteristics of cascades on spatial networks, we generated
rewired RGGs as follows. Starting with the original spatial network,
we rewire an arbitrarily chosen end of each link to a randomly
chosen node in the network with probability $p$. During this
process, we ensure that no self-loops or multiple edges are
generated, by rejecting any rewiring step that leads to these
undesired features.

\subsection*{4. Empirical Network}
 As a realistic testbed on which to validate our results, we use the UCTE European power transmission network from the year 2002 \cite{Bialek05,Bialek13,newdata}, which we will henceforth refer to simply as the UCTE network. This network is spread over a geographic area that comprises 18 countries, and consists of $N = 1254$ buses which constitute the nodes for our
 purposes. The average degree of the network is $\langle k \rangle = 2.89$.

\section*{Results}

\subsection*{1.~Load landscapes in RGGs}

\begin{figure}[!ht]
\begin{center}
\includegraphics[width=5in]{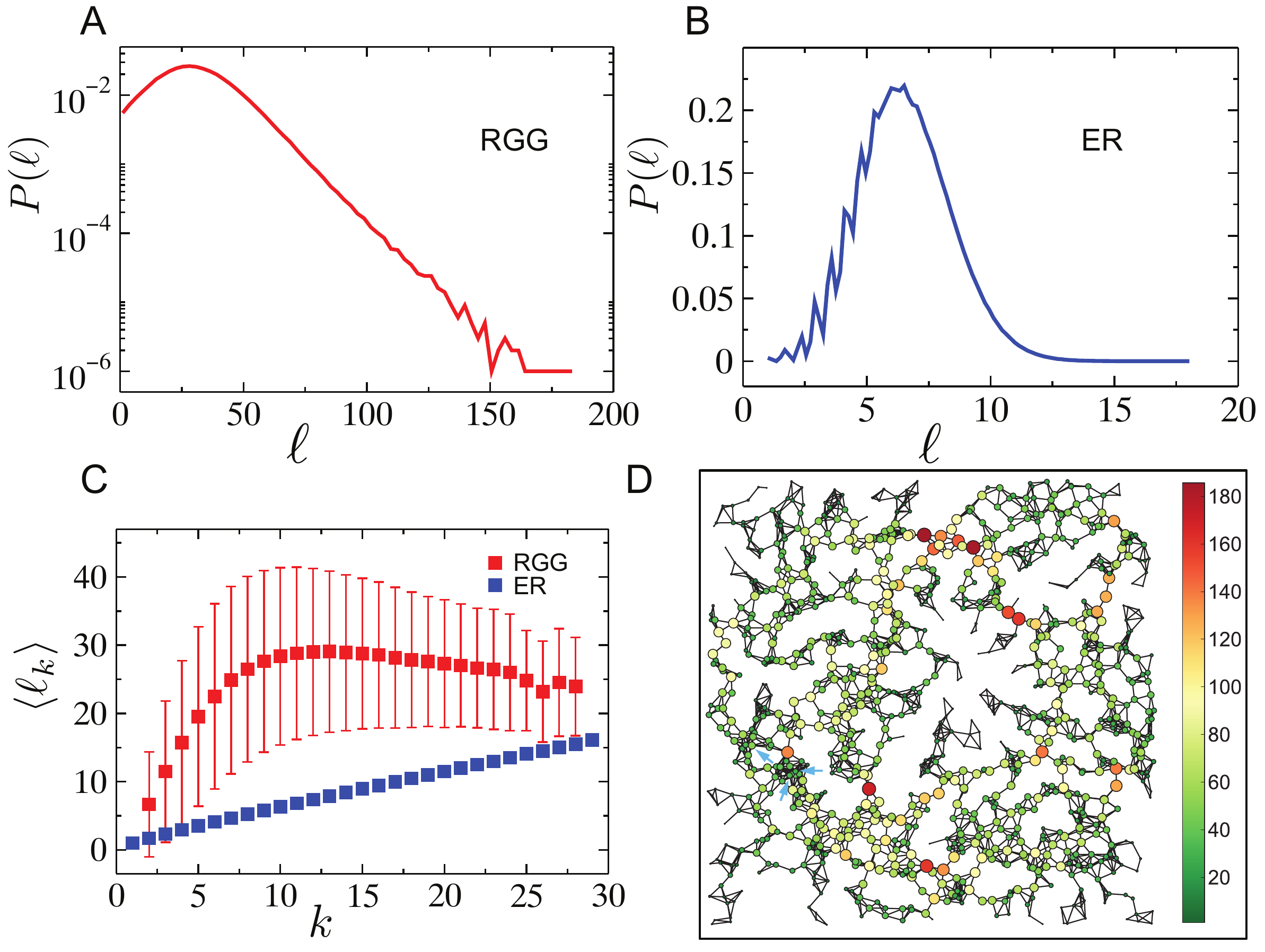}
\end{center}
\caption{
{\bf Vertex load profiles in RGGs and ER networks.} Calculated on (A) random geometric graphs and (B) Erd\H{o}s-R\'enyi random grahs, composed of $N=1500$ nodes with $\langle k \rangle=10$ and averaged over $2000$ network realizations. (C) Positive correlations are shown in the case of ER graphs, while these correlations seem to disappear in RGGs for degree classes higher than the average degree of the network ensemble. Data were averaged over more than $3000$ network realizations for networks of $N=1000$ and $\langle k \rangle=10$. The fluctuating tail of the red curve originates from the lack of sufficient number of samples in the specific degree classes. The error bars correspond to one standard deviations. (D) A single network realization showing the vertex loads. Note, that the node with the highest connections (blue arrows indicate the 3 highest degree nodes) does not carry the highest load in the network (loads are color coded, and node sizes are proportional to loads).}
\label{fig:1}
\end{figure}

We begin by analyzing the vertex load distributions in RGGs and
comparing them to those in ER graphs with the same average degree,
the latter playing the role of a null-model where spatial
constraints are absent. Both RGGs and ER graphs are characterized by
homogeneous (Poissonian) degree distributions \cite{Herrmann_PRE03}.
In addition, RGGs exhibit a high clustering coefficient
\cite{Dall02}, resulting from the spatial dependance of
the connectivity and the transitivity of spatial relationships. Path
lengths on RGGs scale with the network size $N$ in contrast to the
$\log N$ scaling found in ER graphs.
Given these differences, we expect that the vertex load distribution
for RGGs would also differ significantly from that of ER graphs. Indeed, as shown in Figs.~\ref{fig:1}A and
\ref{fig:1}B respectively, the vertex load distribution for RGGs has
an exponentially decaying tail with a decay constant $\approxeq
0.083$, while the distribution for ER graphs is best-fitted by a
Gaussian distribution (parameters in caption). For identical average
degrees, the mean vertex-load in RGGs,  ($\langle \ell
\rangle=32.54$), is almost six times as large as that for ER graphs.
Figure ~\ref{fig:1}C shows the average vertex-load conditioned on
the vertex-degree, as a function of the degree. Again, in contrast
to the case of ER graphs, the plot for RGGs does not display an
unambiguously positive correlation of vertex-load with degree over
the entire degree range. The vertex-loads are strongly correlated
with degrees up until a value close to the average degree, after
which they show a subtle decline. A visualization of the network
(Fig.~\ref{fig:1}D) makes it clear that the nodes with the highest
loads do not have degrees anywhere as high as the largest degree in
the network.

For a network where connections are spatially constrained, we
intuitively expect that a high load on a node is indicative of a
high load in its neighborhood. To substantiate this, we investigate
the spatial correlation between vertex loads on the network.
Specifically, we measure the correlation between vertex loads as a
function of the distance separating them, by systematically
obtaining all pairs of nodes $(i,j)$ separated by a distance that
lies within $(r-\Delta r,r + \Delta r)$, and computing the Pearson
correlation coefficient between these pairs:
\begin{equation}
C_L(r) = \frac{\sum_{i,j|r_{ij} \in (r-\Delta_r/2,r+\Delta_r/2)}(\ell_i - \langle \ell_i \rangle)(\ell_j - \langle \ell_j \rangle)}{\sigma_{i} \sigma_{j}}
\label{pearsoncorrelation}
\end{equation}

\begin{figure}[t]
\begin{center}
\includegraphics[width=5in]{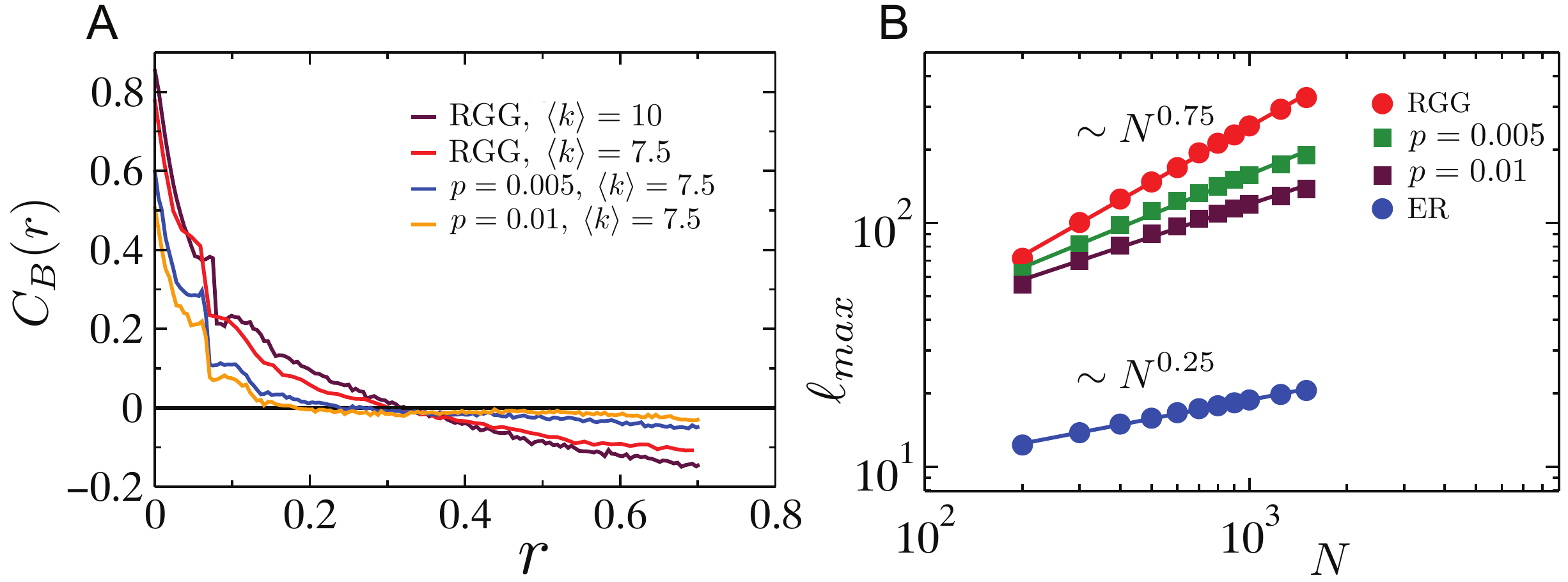}
\end{center}
\caption{{\bf Correlations of vertex loads as a function of distance between vertices and extreme value characteristics of loads.} (A) Load and distance correlations in RGG. Pearson correlation coefficient as function of distance $r$ measured between two arbitrary nodes. Data were averaged over $100$ network realizations for networks having $500$ nodes. (B) System size dependence of the maximum vertex load in networks with $\langle k\rangle=6$. Data were averaged over $2000$ network realizations. The parameter $p$ corresponds to the rewiring probability for links in the RGG.}
\label{fig:2}
\end{figure}

Figure~\ref{fig:2}A shows the Pearson correlation coefficient
between loads at a distance $r$ apart from each other. $150$ evenly
spaced values of $r$ were considered within the complete range $(0,
\sqrt{2}/2)$, with $\Delta_r = \frac{\sqrt{2}/2}{150}$. The
resulting picture shows that loads are positively correlated for
nodes within a distance $r = R$ where $R$ is the connection radius,
while just beyond this value the correlation sharply drops and
continues to decrease monotonically thereafter, reaching slightly
negative values at very large separations. The picture obtained for
networks with different average degrees is qualitatively similar,
and does not change significantly for rewired RGGs generated using
small values of the rewiring parameter. It is worth mentioning that
although the spatial correlation captured by the Pearson correlation
coefficient indicates vertex loads being correlated only within a
short distance, it does not preclude the existence of lower
dimensional correlated structures - such as a 1D backbone formed by
vertices with high loads \cite{Marek10} -within the network. To
conclude this study of the load profiles, we analyze the extreme
value scaling of the load distribution with network size $N$, a
quantity of significance in determining the effective throughput of
the network \cite{SameetBottleneck07}. As shown in
Fig.~\ref{fig:2}B, the maximum vertex load on RGGs scales as a power
law with $N$, with an exponent of $0.75$. This is a much faster
growth in comparison to the scaling,$\sim N^{0.25}$ found for ER
graphs. Rewiring the links of the RGG with increasing probability
$p$, gradually but systematically lowers the loads, and their
scaling. (The scaling exponents found for $p = 0.005$ and $p = 0.01$
are $0.545$ and $0.44$ respectively.)

\subsection*{2.~Cascades of overload failures}

Next, we simulate cascading failures on a network triggered either by random or targeted removals of nodes, and quantify the resilience of the network to such failures. The model used (see Methods) is identical to that used in earlier studies \cite{Motter04, MotterLai02, Asztalos12}, and is parametrized by a single tolerance parameter $\alpha$ which quantifies the excess load bearing capacity of a node. Following the notation introduced in \cite{MotterLai02}, the resilience of a network is quantified in terms of the fractional size of the surviving largest connected component after the cascade ends: $G=N'/N$, where $N'$ is the number of nodes belonging to the largest network component after the cascade and $N$ is the undamaged (connected) network size.

\begin{figure}[!ht]
\begin{center}
\includegraphics[width=5in]{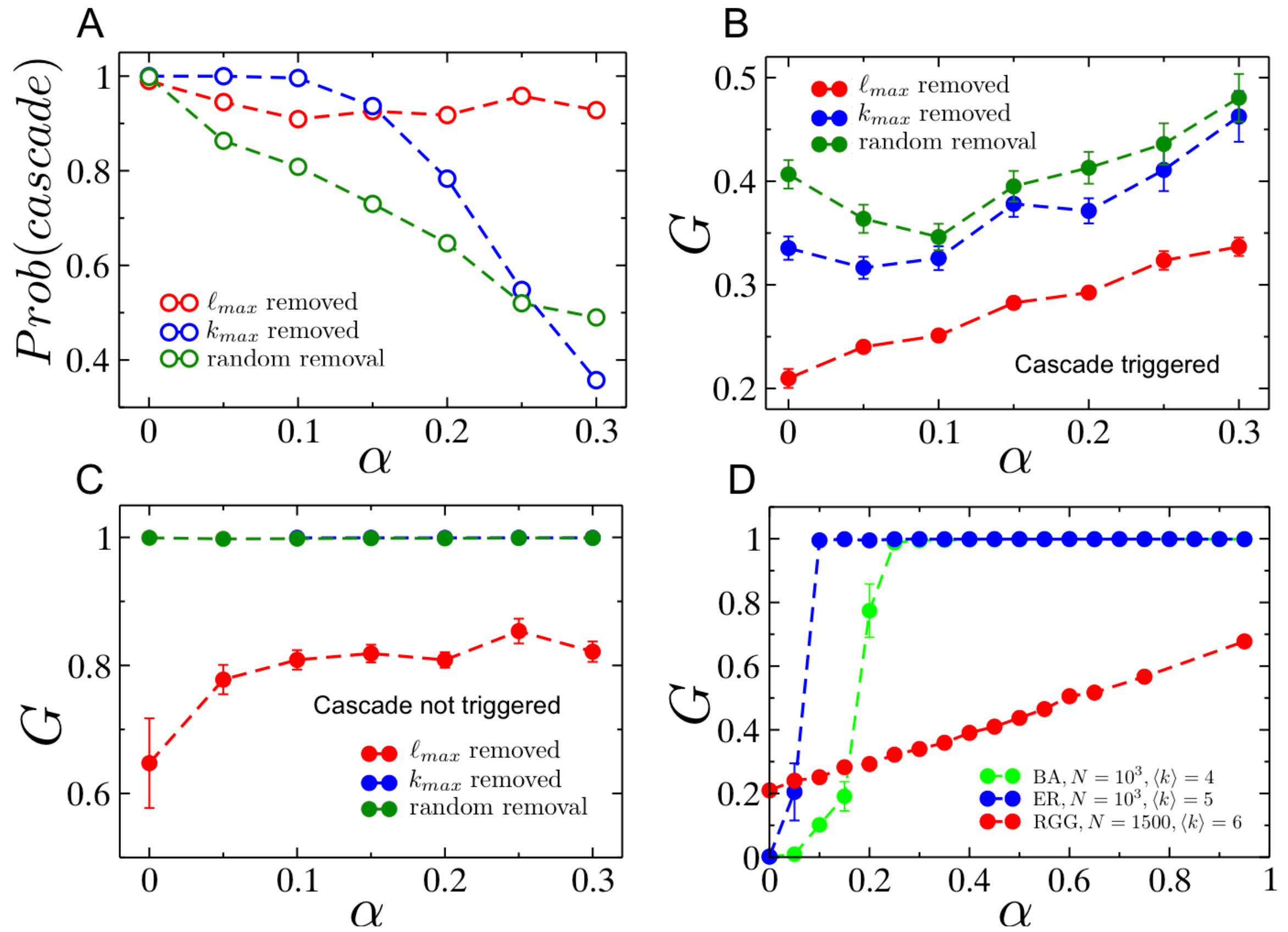}
\end{center}
\caption{{\bf Cascades triggered by targeted and random removals.} (A) Probability that a single node removal will trigger a cascade as function of the tolerance parameter. (B) The ratio $G$ of the size of the largest surviving network component to the initial network size, as function of $\alpha$, the tolerance parameter when the initial failure triggers a cascade. (C) Similar to (B), except for the case where the initial failure does not trigger a cascade.  In all (A), (B) and (C) subplots the red curve corresponds to the case when the triggered node is the node with the highest load, the blue curve to the case when the triggered node is the most connected node in the network and the green curve shows the case when the triggered node was chosen randomly. Network parameters are $N=1500$, $\langle k \rangle=6.0$, while the data was averaged over $500$ network realizations. Error bars correspond to the standard error of the mean. (D) $G$ as a function of tolerance parameter, unconditioned on whether or not a cascade was triggered for RGGs, SF networks and ER networks.}
\label{fig:3}
\end{figure}

Figure~\ref{fig:3}A shows the probability that a cascade ensues after an initial node removal. As seen, irrespective of the tolerance parameter, cascades triggered by the removal of the node with the highest load in the network leave behind the largest damage when compared with those resulting from removal of the highest degree node or a random node. Figures~\ref{fig:3}B,C show the fractional size of the surviving giant component $G$ as a function of the tolerance parameter in the presence and the absence of a cascade. Once again, the damage done is the worst for the case where the initial node removed is the one with the maximal load, even in the case where no cascade is triggered, suggesting that vertices with the highest loads are those responsible for bridging together distinct connected components and ensuring the structural integrity of the network. Finally, Fig.~\ref{fig:3}D compares the damage done due to cascading failures on RGGs with the damage in SF and ER networks, all having the same average degree. Clearly, while increasing excess capacity does lead to an increase, on average, of the surviving giant component, the growth is profoundly slower for RGGs than for the spatially unconstrained networks. Henceforth, as we further investigate more detailed characteristics of cascades, we restrict our studies to cascades triggered by the removal of the vertex with the highest load, since the damage done to the network is the most severe in this case.

\begin{figure}[!ht]
\begin{center}
\includegraphics[width=5in]{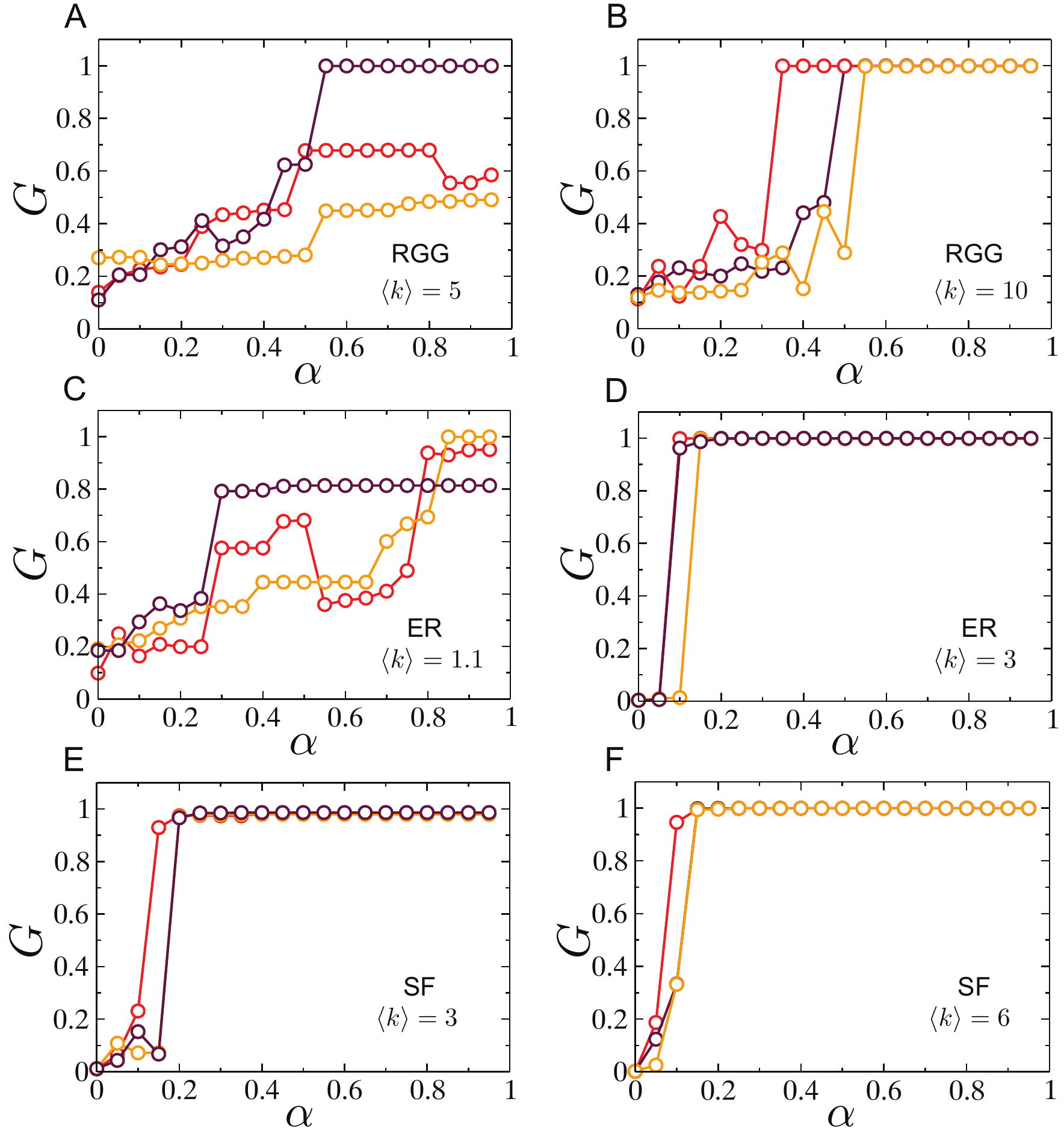}
\end{center}
\caption{{\bf Cascades on single network realizations.} Simulations were performed on networks of size $N=1300$. Fractional size of surviving giant component as a function of $\alpha$ for (A),(B) RGGs, (C),(D) ER networks and (E),(F) SF networks.}
\label{fig:4}
\end{figure}

As shown above, increased capacity allocation results in a monotonic increase in the average surviving giant component size, where the averaging is done over an ensemble of network realizations. If such a monotonic increase was also obtained for individual network instances, then increased capacity allocation, although only weakly effective, would at least be a justifiable preventative measure against cascades.  Figures~\ref{fig:4} A, B show the size of the surviving giant component $G$ as a function of the tolerance parameter $\alpha$ for three individual instances of RGGs, for different respective average degrees. As is clearly seen, the variation in $G$ is far from monotonic for a single network instance, and differs significantly across instances. Thus, the trend observed by averaging a macroscopic quantity, $G$, over an ensemble of RGG networks (as was the case in Fig.~\ref{fig:3}) provides little indication of the true behavior of the same quantity for an individual network instance. This behavior persists even if the network size is increased (not shown). Such lack of self-averaging has been observed previously in fragmentation processes on lattices, to which cascades bear a close resemblance \cite{Krapivsky2000}. In contrast, results of cascades on single instances of ER and SF networks, shown in Figs.~\ref{fig:4}B,C, are consistent with those obtained by averaging $G$ over respective network ensembles.

\begin{figure}[!ht]
\begin{center}
\includegraphics[width=5in]{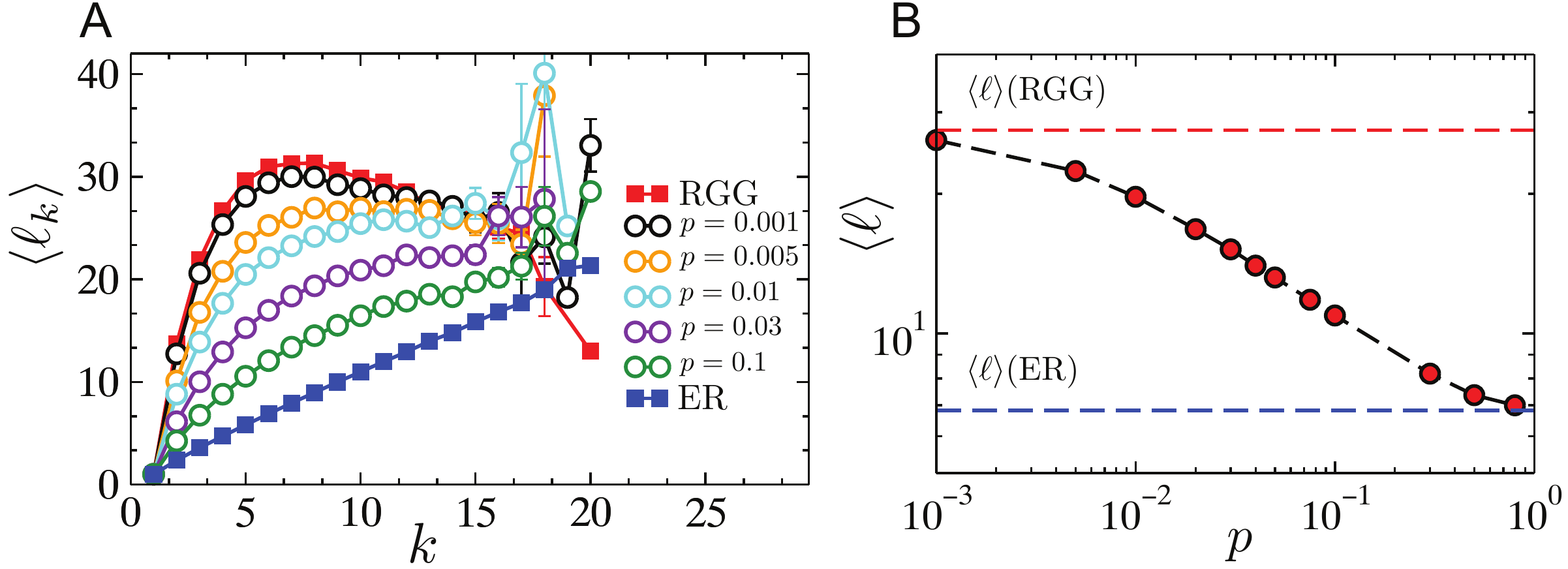}
\end{center}
\caption{{\bf Effect of rewiring in RGGs.} (A) Transition in network structure from an RGG towards an ER network through the process of rewiring. Multiple rewired versions are shown together with the two extreme cases. (B) Average vertex load in RGG, ER and rewired versions of RGG as function of the fraction of rewired links $p$. Network parameters are: $N=500$, $\langle k \rangle=6.0$. Data were averaged over $500$ network realizations.}
\label{fig:5}
\end{figure}

\begin{figure}[!ht]
\begin{center}
\includegraphics[width=5in]{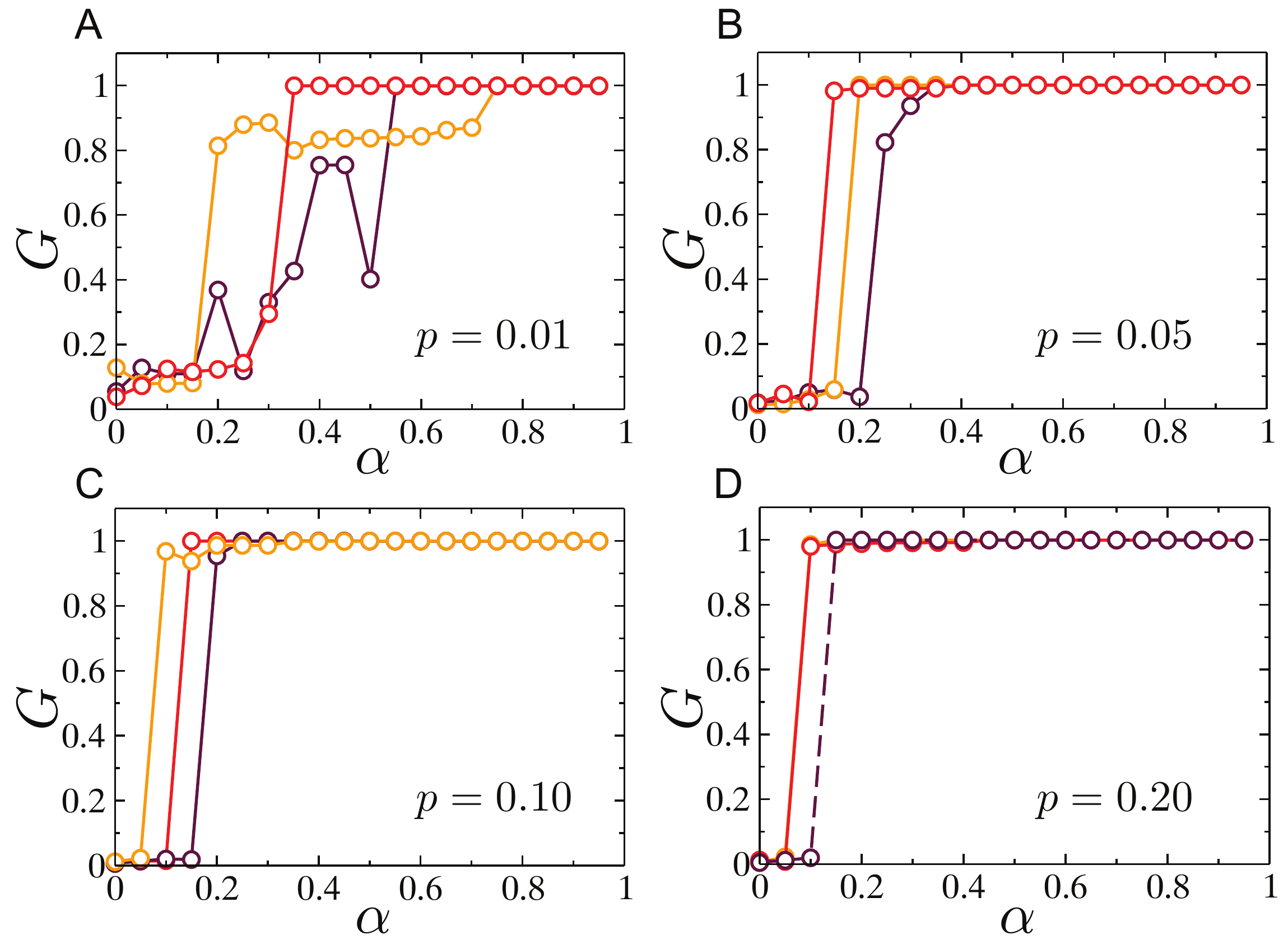}
\end{center}
\caption{{\bf Effect of rewiring on cascades in RGGs.} Cascades were triggered by the removal of the highest load. Simulations were performed on networks of size $N=1300$ with $\langle  k \rangle=5.0$. As $p$ is increased the lack of self-averaging manifested by the non-monotonicities in the curves for $G$ versus $\alpha$ disappears.}
\label{fig:6}
\end{figure}

Presumably, this lack of self-averaging is a feature that results
from the embedding of the network in two-dimensions (with no
shortcuts). To conclusively validate this argument, we study how the
presence of a few spatially unconstrained links affects the
surviving giant component size, since the addition of such links has
the effect of increasing the underlying dimensionality of the space
in which the network is embedded. Specifically, for each link,  we
rewire with probability $p$ one end of the link with a randomly
chosen node in the network, without allowing self-loops or multiple
edges to form. Similar constructions have been used before in
\cite{SmallWorld98,RGG_rewired09,Qiming_PRE2008}. By varying $p$
between $0$ and $1$, we can interpolate between RGGs and ER graphs,
as is confirmed by the results shown in Fig.~\ref{fig:5}, where both
the degree-conditioned average load and the average load undergo a
smooth crossover from the results expected for RGGs to those
expected for ER graphs. Figure~\ref{fig:6} shows that even with as
low as $5\%$ of the links of the RGG rewired, the non-monotonicity
in $G$ versus $\alpha$ completely disappears. The interval of $p$
within which the crossover takes place contains values larger than
the theoretical estimate of $p^* \sim 1/(\langle k \rangle N/2)$
\cite{Newman99} at which the small-world crossover occurs, likely a
finite-size-effect due to the small system sizes considered here.
Thus, from a theoretical network-design point of view, the incorporation of a
few long-range links would be a simple step in stabilizing flows and managing cascades,
since it results in a more predictable relationship between surviving-component
size and excess capacity. However, in practical
situations the cost of adding such long-range links could be prohibitive, and therefore
may not constitute a feasible solution for controlling the grid.

We also studied how length dependent link-conductances affected our results. Specifically,
we assumed that $C_{ij} = A_{ij}/d_{ij}$ for a link connecting nodes $i$ and $j$ where $d_{ij}$ denotes the Euclidean distance between them, and performed
simulations to study the dependence of the surviving giant component size $G$ as a function
of the tolerance parameter $\alpha$ (analogous to results in Fig.~\ref{fig:4}~A,B), and to investigate
the effect of rewiring links to create a few long-range connections in the network (similar to the results in Fig.~\ref{fig:6}). For both
cases, we found no significant quantitative difference in the results for the case where conductances
were length-dependent. In particular, the non-self-averaging nature of cascades manifested itself
in exactly the same manner as is demonstrated in Supplementary Figs. S1 and S2.

\begin{figure}[!ht]
\begin{center}
\includegraphics[width=5in]{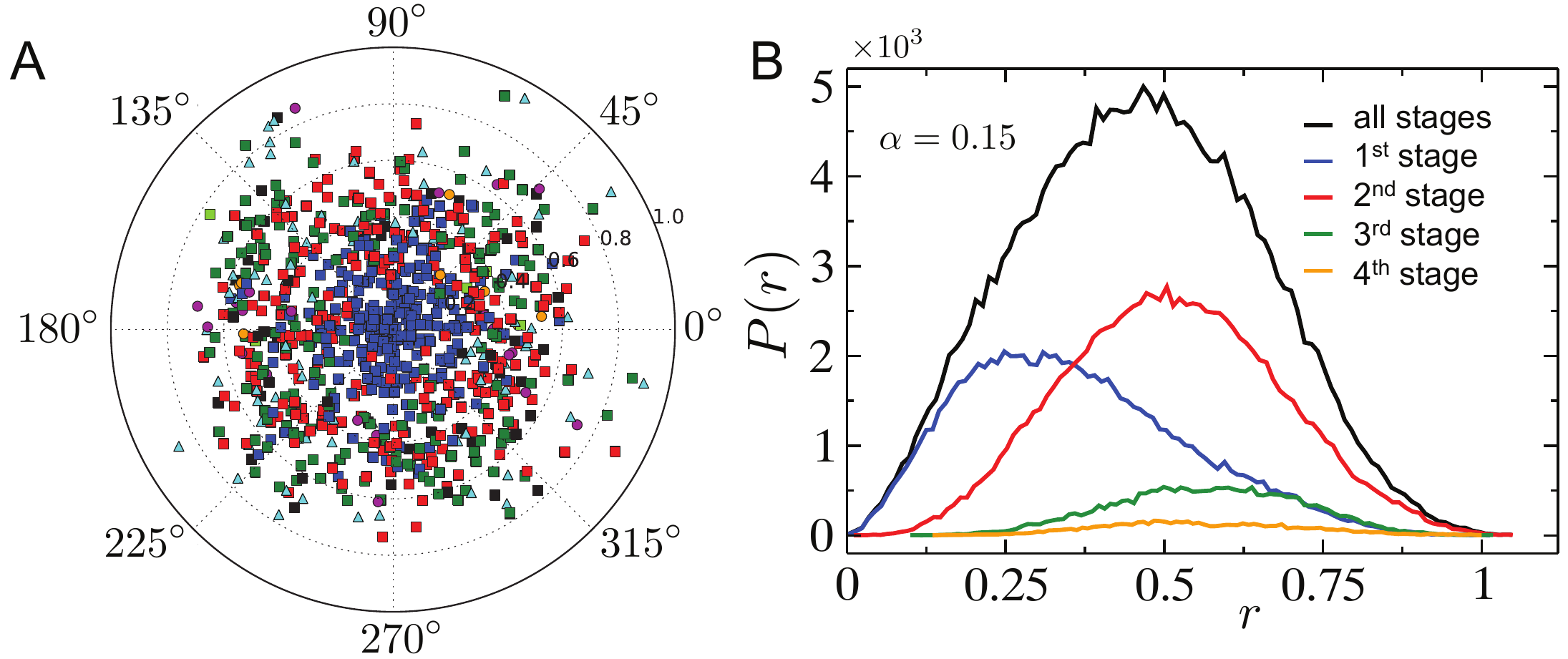}
\end{center}
\caption{{\bf Location of overloaded nodes.} (A) Position (distance and angle) of failed nodes relative to the initially removed one, here the highest load in the network. Different colors correspond to different iterations of the cascade: blue squares (1st), red squares (2nd), green squares (3rd), light blue triangles (4th), black squares (5th), magenta circles (6th), orange circles (7th), light green squares (8th), yellow triangles (9th). Network parameters are the same as in Fig.~\ref{fig:1}, while each data point is the average value of $540$ cascades. (B) Probability density function of the distance $r$ from the cascade-triggering node for nodes that fail in the course of a cascade. }
\label{fig:7}
\end{figure}

As a next step in understanding the nature of spatial cascades, we measure the spatial correlations between nodes that fail in successive stages of the cascade. Here, a single {\it stage} refers to a round of calculating vertex loads, and removing those nodes whose load exceeds their respective capacity. Figure.~\ref{fig:7}A shows the {\it average} location of failing nodes per stage of the cascade, relative to the node that triggers the cascade. The most significant feature observed here, as well as in the distribution of distance (from the cascade-triggering node) for failing nodes in each cascade stage (Fig.~\ref{fig:7}B) is the separation between the most likely locations for nodes removed in the first and second stages.  In subsequent stages, the distribution of the location of failing nodes gets progressively more uniform. In general, as seen from our simulations, cascades last for only a few stages  (the longest found in the systems studied here was $11$ stages) with most of the damage occurring by the second stage, and then declining rapidly. The stage-wise distributions in Fig.~\ref{fig:7}B were obtained by aggregating all nodes removed in a particular stage and belonging to a particular distance bin over $540$ distinct cascades, and normalizing them by the total number of nodes removed over the distinct cascades. Thus, declining contribution of later stages is due to a combination of two factors: the reduction in the number of nodes removed during later stages, and the decrease in the probability of the cascade surviving up to that stage. The all-stage distribution was generated in a similar fashion as the stage-wise distribution, but disregarding the stages associated with the nodes.

\begin{figure}[!ht]
\begin{center}
\includegraphics[width=4in]{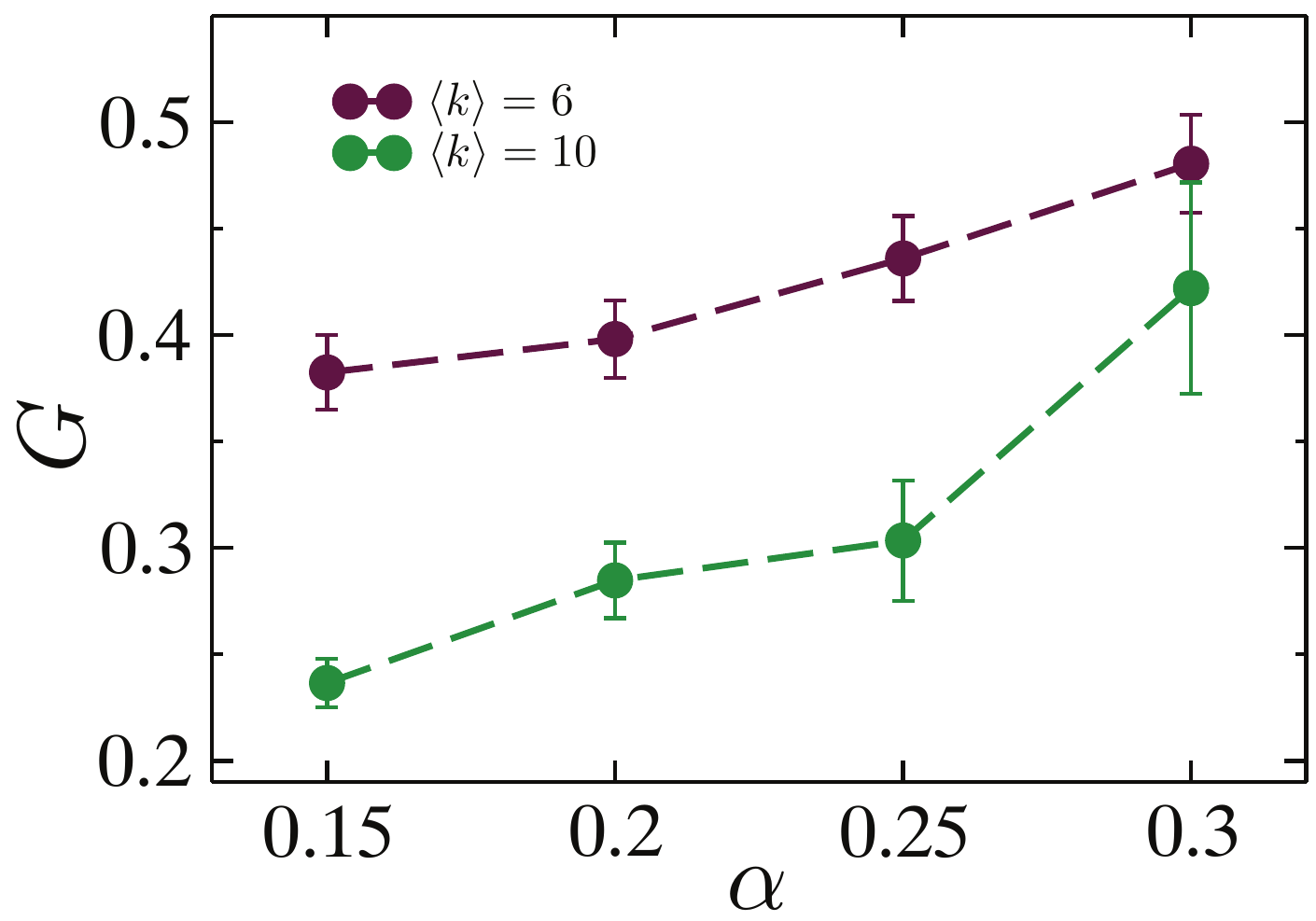}
\end{center}
\caption{{\bf The effect of average degree upon cascading failures.} Fraction of the largest surviving network component following cascading failures $(G)$ triggered by the removal of a single, randomly chosen node as function of $\alpha$ tolerance parameter. The two curves correspond to two ensembles of random geometric graphs, one with $\langle k\rangle=6$ (maroon) and one with $\langle k\rangle=10$ (green). Data were obtained for RGGs of size $(N=1500)$, averaged over more than $400$ network realizations. The error bars correspond to the standard error of the mean.}
\label{fig:8}
\end{figure}

Finally in this section, we study the effect of average degree of RGGs on their resilience to cascading failures. Figure~\ref{fig:8} compares the fractional size of the largest connected component as a function of $\alpha$ for networks with average degree $\langle k \rangle=6$ and $\langle k \rangle=10$. Surprisingly, the damage caused by cascading failures is far more severe for the more well connected of the two network ensembles. Although, for other dynamical processes such as epidemic spreading and diffusion it is intuitively obvious that more connections lead to more spread, here we would expect that the presence of more paths between any source-sink pair on a denser network would lead to more effective load balancing, and therefore weaker cascading failures.
However, while increasing the average degree does cause loads for each node to be lower and more balanced initially, the excess capacity allocation in proportion to these lower and more uniform loads, makes the network ill-equipped to handle variations in load resulting from the initial node removal. As a result, cascades cause more damage for a denser RGG than a sparser one.
In contrast, as is well known, denser RGGs are structurally more resilient to random (non-cascading) failures occurring in the network, since the giant component undergoes a transition in size at $f_c =1 - \frac{\langle k_c \rangle}{\langle k \rangle}$ \cite{Diaz01,Kong10} where $f_c$ is the critical fraction of randomly removed nodes from a RGG, and $\langle k_c \rangle$ is an embedding-dimension dependent constant taking the value $4.52$ \cite{Dall02} for two dimensions.

\subsection*{3.~Cascade mitigation strategies}

\begin{figure}[t]
\begin{center}
\includegraphics[width=5in]{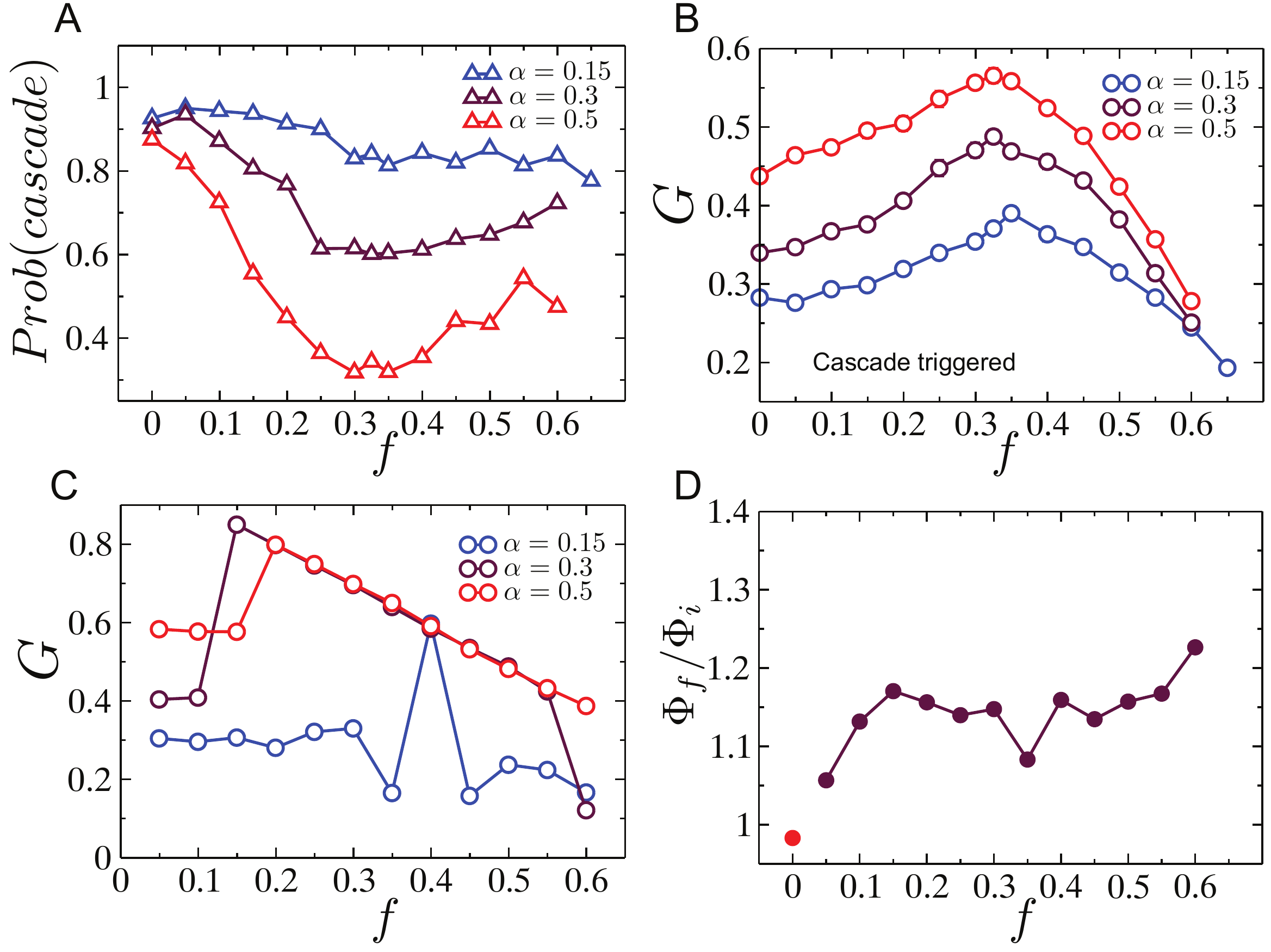}
\end{center}
\caption{{\bf Preemptive node removal in RGGs.} (A) Probability that a cascade occurs after removal of the node with highest load, despite a fraction $f$ of nodes being preemptively removed immediately after the initial trigger. (B) Fractional size of the largest surviving network component $G$ as a function of preemptively removed fraction $f$, when there is a cascade. (C) Fractional size of the largest surviving network component $G$ as a function of preemptively removed fraction $f$ for a single network instance for different values of the tolerance parameter $\alpha$. (D) The ratio of the throughput (defined in text) of the surviving giant component and the throughput of the original network as a function of the altruist node fraction. The red circle corresponds to the case when there no nodes are preemptively removed. Network parameters are: $N=1500$, $\langle k \rangle=6.0.$}
\label{fig:9}
\end{figure}

Next, we study two cascade mitigation strategies and evaluate their effectiveness. We begin by analyzing the preemptive node removal strategy proposed by Motter \cite{Motter04}. Intuitively, this method aims to utilize node removal in such a way that the two competing objectives of reducing the load on the network, and keeping the network connected, are balanced. Specifically, the method involves removing a fraction $f$ of the lowest load nodes after the initial node failure. This method was motivated by studies on scale-free networks where the load distribution is heavy-tailed implying that a significant fraction of nodes despite contributing to the total load on the network by acting as sources of current/packets, only frugally participate in the carrying of loads generated by other source-sinks pairs, due to their low betweenness. The load distribution for RGGs however, is comparatively much narrower, and we would therefore expect that preemptive node removal would not yield significant success. The results of investigating the efficacy of preemptive node removal as a cascade mitigation strategy are presented in Fig.~\ref{fig:9}.  As shown in Fig.~\ref{fig:9}A, the probability of a cascade occurring decreases (with increasing $f$) until it reaches a minimum, and beyond which, it increases again. A similar profile is also observed for the ensemble averaged values of the fractional size of the giant surviving component $G$ as a function of $f$. Both plots show however, that even at the optimal $f$, and for as large as $50\%$ additional capacity (i.e. $\alpha = 0.5$), the gains obtained are weak. Furthermore, as a consequence of the lack of self-averaging, individual network instances show profiles that are highly variable and showing little resemblance to the ensemble averaged results. Three such examples for individual network instances are shown in Fig.~\ref{fig:9}C. Finally, we study how the throughput in the giant surviving component after a cascade, $\phi_f$, compares to the throughput on the original network, $\phi_i$. The throughput captures  the maximum current that can be injected per source without the network becoming congested. For $\phi$ units of current injected at every source, the network is uncongested if for every node $j$, the inequality, $\phi_j \leq C_j$ holds. Consequently, for the intact network (indicated by subscript $i$), the throughput is $\phi_i = \frac{1}{\max_j \{l_j/C_j\}}$. The throughput can similarly be calculated for the surviving component after a cascade. As shown in Fig.~\ref{fig:9}D, the throughput after the cascade $\phi_f$ is larger than the initial throughput for $f>0$. The increase in throughput is expected since the size of the network is smaller after a cascade, leading to a reduction in loads (due to the $N$ dependance in the definition of loads, see Eq.~\ref{eq:2}) and thereby an increase in the quantity $\max_j \{l_j/C_j\}$. For the case where $f=0$, although the ensemble average of the ratio $\phi_f/\phi_i$ is smaller than one ($\approx 0.98$), in most individual instances the ratio is exactly one. In these cases, the throughput after the cascade is determined by a node whose connectivity before and after the cascade is $k=1$. Such a {\it dangling end} has initial load equal to $1$ which remains unchanged after the cascade as well i.e. the reduction in the number of sources and sinks in the system has no effect on its load, unlike for other nodes which have higher connectivity. Therefore the value of $l_j/C_j$ after the cascade for such a node often ends up being the highest among all nodes, and by definition results in the throughput after the cascade being identical to that before the cascade i.e. $(1 + \alpha)$. When $f >0$, such dangling ends are removed as part of the preemptive node removal process, and all surviving nodes end up experiencing a reduction in load due to the reduced size of the surviving giant component. As a result, the final throughput is higher than the initial throughput, resulting in $\phi_f/\phi_i$ being greater than one.

\begin{figure}[!ht]
\begin{center}
\includegraphics[width=5in]{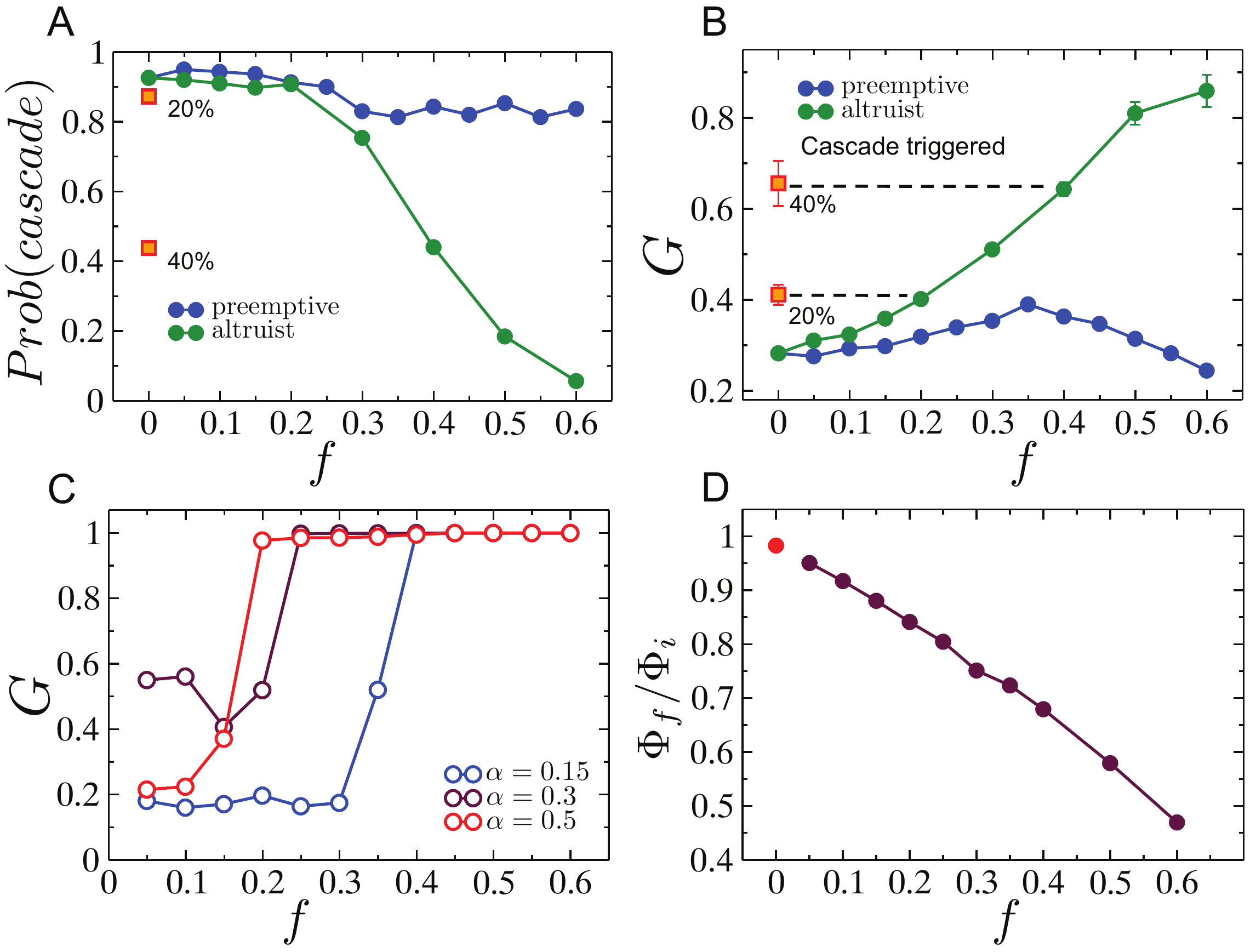}
\end{center}
\caption{{\bf Increasing the resilience of the network by introducing altruist nodes.} (A) Probability that a cascade is triggered for an altruist/preemptively removed fraction $f$. The orange squares indicated the probability of cascade when no nodes (other than the initial cascade-triggering node) are removed, but when the current per source is reduced by $20\%$  (upper square) or $80\%$ (lower square) immediately after the initial node removal.
(B) The fractional size of the surviving giant component $G$ when a cascade is triggered, as a function of the altruist/preemptively-removed node fraction. Also shown are the results when the current per source is reduced by $20\%$  (upper square) or $80\%$ (lower square) immediately after the initial node removal, which coincide with the $f=0.2$ and $f=0.8$ results respectively for altruistic node removal. (C) Similar to (B), but for the cases where a cascade is not triggered. (D) The ratio of the effective throughput (defined in text) of the surviving giant component and the throughput of the original network as a function of the altruist node fraction. The red circle corresponds to the case when there are no altruist nodes. Network parameters for all these plots are: $N=1500$, $\langle k \rangle=6.0$, $\alpha$=0.15. }
\label{fig:10}
\end{figure}

In view of the observation that the pre-cascade vertex load distribution in the RGG is not highly skewed, we propose a cascade mitigation strategy where rather than reducing the total load on the network by the making a fraction of nodes ``absent" from the network as we did for the preemptive strategy, we assign a random fraction $f$ of nodes to be altruists who cease to act as sources in the event of a node failure, but continue conducting flow between other source-sink pairs. Figure~\ref{fig:10}A shows the drop in the probability of a cascade as a function of the fraction $f$ of altruistic nodes. Clearly, the drop is significant in comparison to that achieved by the preemptive node removal strategy. We also show the results of a third strategy which involves all surviving nodes reducing the net current they inject into the network (per sink) to a fraction $f$ of its original value. We show the results (in red) for the two values of $f = 0.2,0.4$, and note that the probability of a cascade is approximately the same as that obtained when only a fraction $f$ of nodes are fully altruistic (i.e. inject no current into the network). Figures~\ref{fig:10}B and C show comparative plots of the size of the surviving giant component obtained for each of these strategies, conditioned on whether a cascade occurs or not. In both cases, the altruistic strategy, as well as the overall current reduction strategy, show a significant improvement over the preemptive node removal strategy. Understandably, this improvement comes at the cost of the overall throughput in the network. Figure~\ref{fig:10}D shows the {\it effective} throughput in the surviving component $\phi_f$ normalized by the initial throughput $\phi_{i}$ of the intact network, as a function of the altruist fraction $f$.
For a principled comparison of the throughputs before and after the cascade , we define the effective throughput of the surviving giant component as the current per source on the intact network that would yield the same total current as that flowing through the surviving giant component after the cascade. Mathematically, when the number of altruist nodes in the surviving component is $n$, this effective throughput is written as:
\begin{equation}
\phi_f = \frac{1}{\max_j \{l_j/C_j\} }\frac{N-n}{N}
\label{effectivethroughput}
\end{equation}
As seen in Fig.~\ref{fig:10}D, the ratio $\phi_f/\phi_i$ decreases as the altruist fraction is increased, thus indicating that the increased surviving fraction comes at the expense of the throughput of the network.

FIGURE 10

\subsection*{4.~Cascade model on an empirical spatial network: The UCTE network}

\begin{figure}[!ht]
\begin{center}
\includegraphics[width=5in]{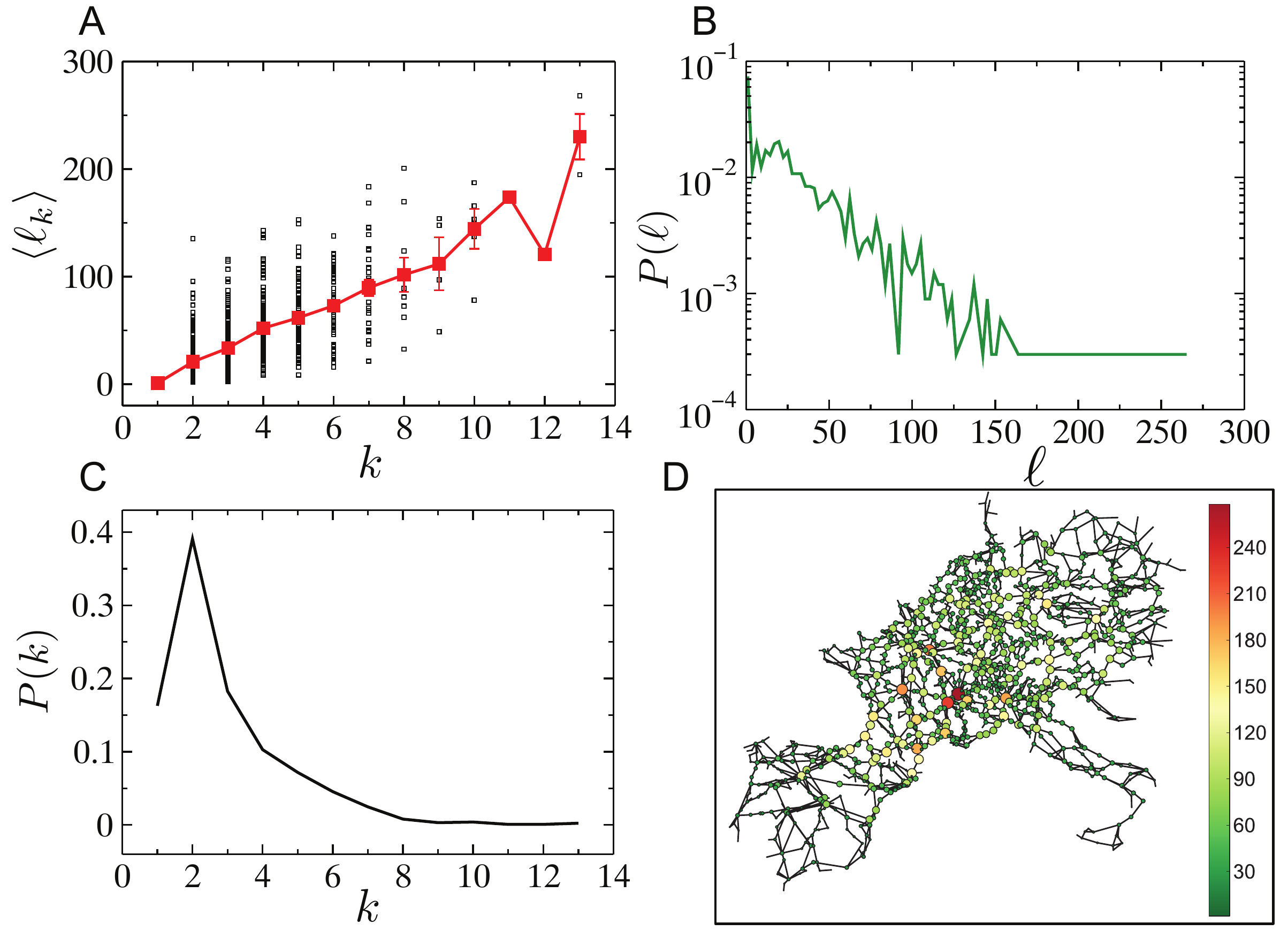}
\end{center}
\caption{{\bf Characteristics of the UCTE network.} (A) The average load across nodes of degree $k$ as a function of $k$. (B) The load distribution on the intact UCTE network. (C) The degree distribution of the UCTE network. (D) A visualization of the UCTE network with loads indicated using both node size and color.}
\label{fig:11}
\end{figure}

Thus far, our studies have been confined to a stylized model of a spatial network, viz. the RGG. We now study the outcomes of the same cascading failure model on the UCTE network, several aspects of which, have been studied elsewhere \cite{Bialek05,Sole07,Verma2013}. The network consists of $N=1254$ transmission stations, with an average degree $\langle k \rangle =2.889$, spanning 18 European countries in 2002. The network is disassortative with an assortativity coefficient of $-0.1$, and with a higher average clustering coefficient than an ER graph ($0.127$). Figure~\ref{fig:11} shows several other properties of this network. The load appears to be positively correlated with the degree (Fig.~\ref{fig:11}A), while the degree and load distributions span a relatively narrow range (Fig.~\ref{fig:11}B,C respectively), as observed also for RGGs.  It is worth noting however, that the variance of loads is significant even for small degree values, which makes it difficult to straightforwardly assess the load bearing responsibility of a node purely from its degree.

\begin{figure}[!ht]
\begin{center}
\includegraphics[width=5in]{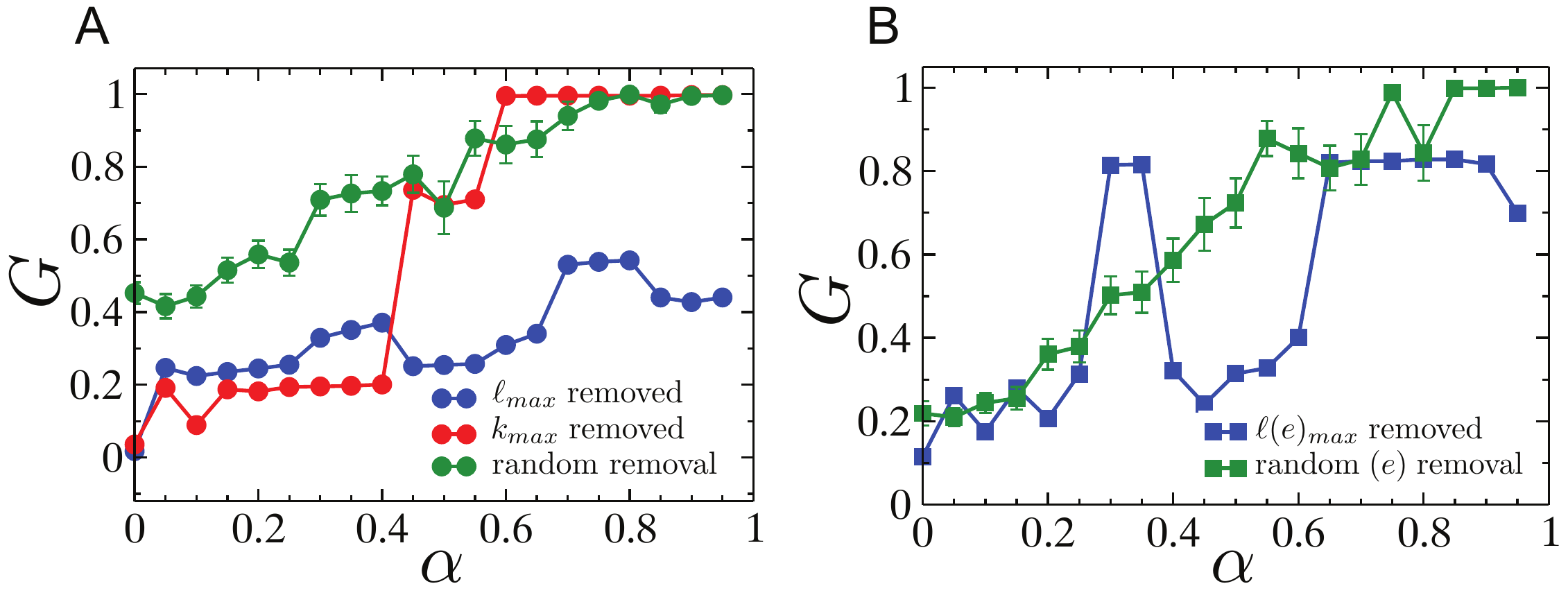}
\end{center}
\caption{{\bf Cascades on the UCTE network.} (A) Cascades triggered by the removal of a single node where the node was chosen using three different criteria i.e. randomly, highest load or highest degree. (B) Cascades triggered by the removal of a single edge where the edge was either chosen randomly or was the one with the highest load. Data obtained for cascade triggered by the random removal of a single node (edge) were averaged over $100$ different scenarios.}
\label{fig:12}
\end{figure}

Figures~\ref{fig:12} A,B show the cascades triggered on the UCTE network by the removal of a a single edge and a single node, respectively. The non-monotonicity observed in $G$ versus $\alpha$ for the model spatial networks is also observed here, thus reinforcing the non-self-averaging nature of spatially constrained networks. In the case of node-removal triggered cascades, removal of the highest-load node results in the worst overall damage, as was also the case for RGGs.

\begin{figure}[!ht]
\begin{center}
\includegraphics[width=3in]{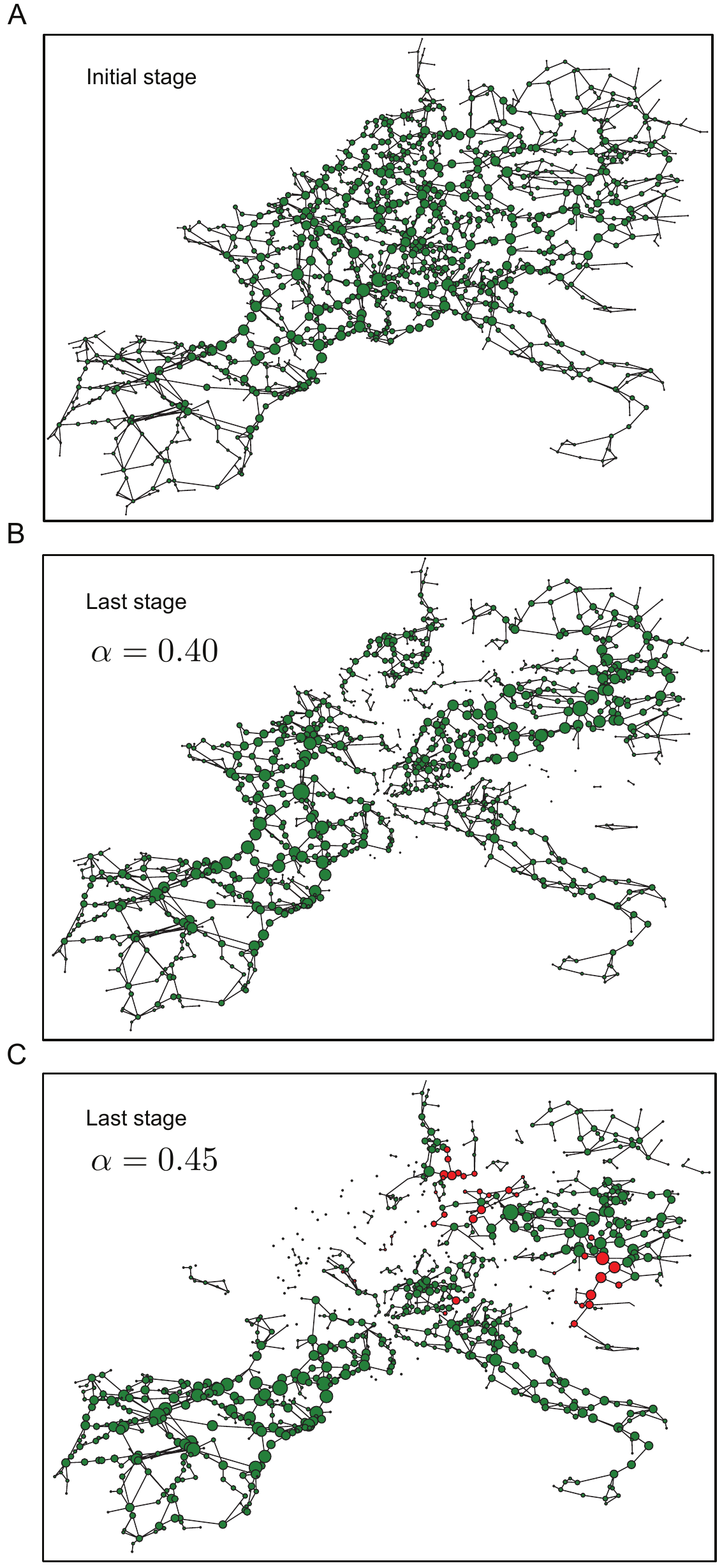}
\end{center}
\caption{{\bf UCTE network snapshots before and after cascades } (A) The intact network with node sizes in proportion to their respective steady-state loads. (B) The network and the loads after a highest-load-removal-triggered cascade has terminated, with the tolerance parameter $\alpha = 0.4$. (C) The network and the loads after a highest-load-removal-triggered cascade has terminated, with the tolerance parameter $\alpha = 0.45$. The red nodes here indicate nodes that were removed in the cascade leading to (B), but survived in the cascade leading to (C).}
\label{fig:13}
\end{figure}

The visualization panels presented in Fig.~\ref{fig:13} provide some intuition on the cause of the observed non-monotonicity in $G$ as the tolerance parameter is increased. Figure~\ref{fig:13} A shows the landscape of loads on the network before the initiation of a cascade where the size of the node is directly proportional to the load on the node. Figure~\ref{fig:13} B shows, the state of the network with tolerance parameter $\alpha = 0.4$ after a cascade initiated by the removal of the highest load, has terminated. Figure~\ref{fig:13} C shows a similar picture for the case where the tolerance parameter is higher, ($\alpha = 0.45$), but where the eventual damage is greater (i.e. $G$ is smaller than the value obtained for Fig.\ref{fig:13} B). In this last panel, the network consists of several nodes, indicated in red, that had been removed in the course of the cascade depicted in Fig.\ref{fig:13} B, but are now intact as a consequence of the increased tolerance. However, counter-intuitively, the survival of these nodes result in wider load imbalances, resulting in a larger overall number of failures and a smaller surviving giant component. Thus, to some degree, the nodes shown in red, behave like fuses which if removed in the course of a cascade, end up saving a larger part of the network from failure. Dynamic visualizations of the progression of the cascades resulting in the final states shown in Figs.\ref{fig:13}B,C are provided in Supplementary Movies S1 and S2, respectively. A feature that becomes apparent in these dynamic visualizations is the non-local nature of the progression of the cascade. As pointed out in \cite{Zussman2011} such non-local progression is commonly observed in real cascade situations, and is a feature which can be reproduced by a more realistic DC power flow model, but not by simpler epidemic or percolation based models. Thus it is worth noting that the model presented in this work, despite being simpler than the DC power flow model used in \cite{Zussman2011}, can nevertheless capture a distinctive attribute of real cascade progression.

\begin{figure}[!ht]
\begin{center}
\includegraphics[width=5in]{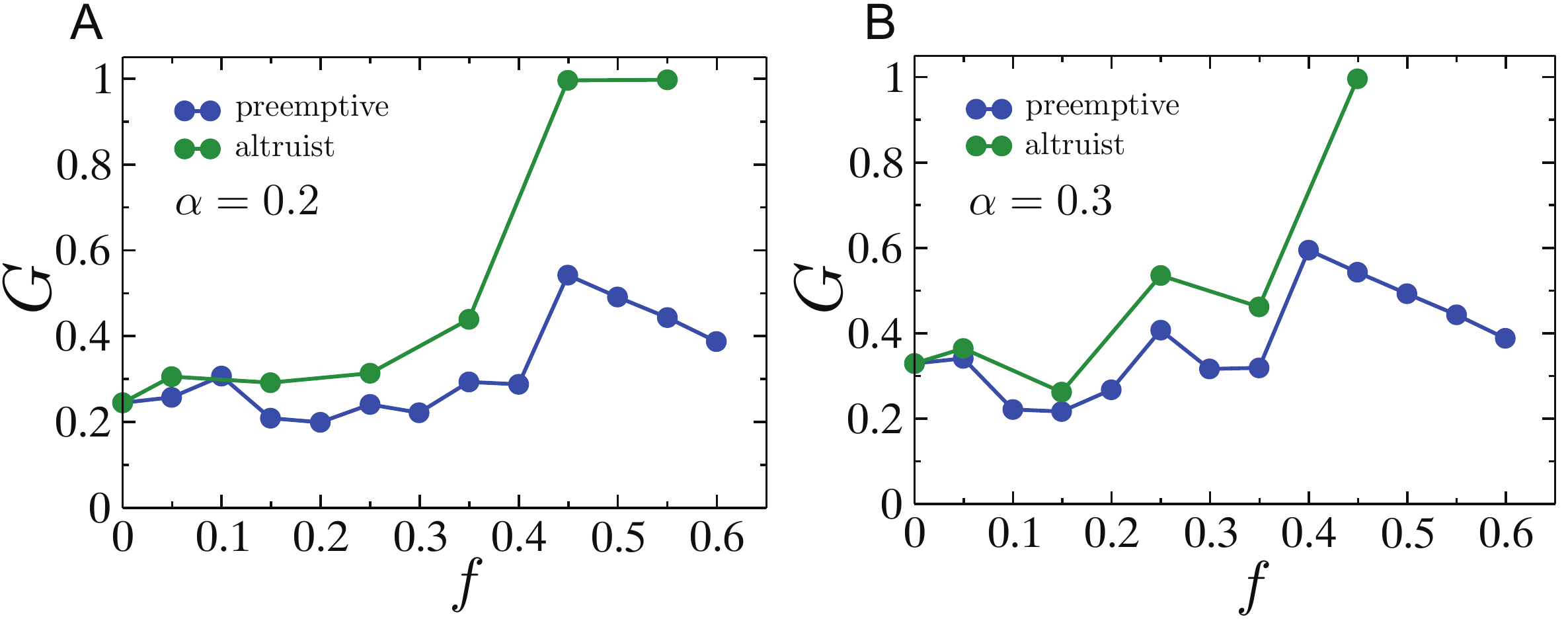}
\end{center}
\caption{{\bf Cascade mitigation on the UCTE network.} Comparison between the preemptive and altruistic node removal strategies on the UCTE network with tolerance parameter (A) $\alpha = 0.2$ and (B) $\alpha = 0.3$.}
\label{fig:14}
\end{figure}

Next, we compare the two cascade mitigation strategies, viz. preemptive node removal and assignment of altruistic nodes, for cascades initiated by highest load removal on the UCTE network. As Fig.~\ref{fig:14}~A and B show, the altruistic strategy generally results in a larger surviving giant component after the cascade, than in the case when preemptive node removal is employed. It is also worth noting that non-monotonicities due to the lack of self-averaging in the cascade process, manifest themselves in these plots as well.

\begin{figure}[!ht]
\begin{center}
\includegraphics[width=5in]{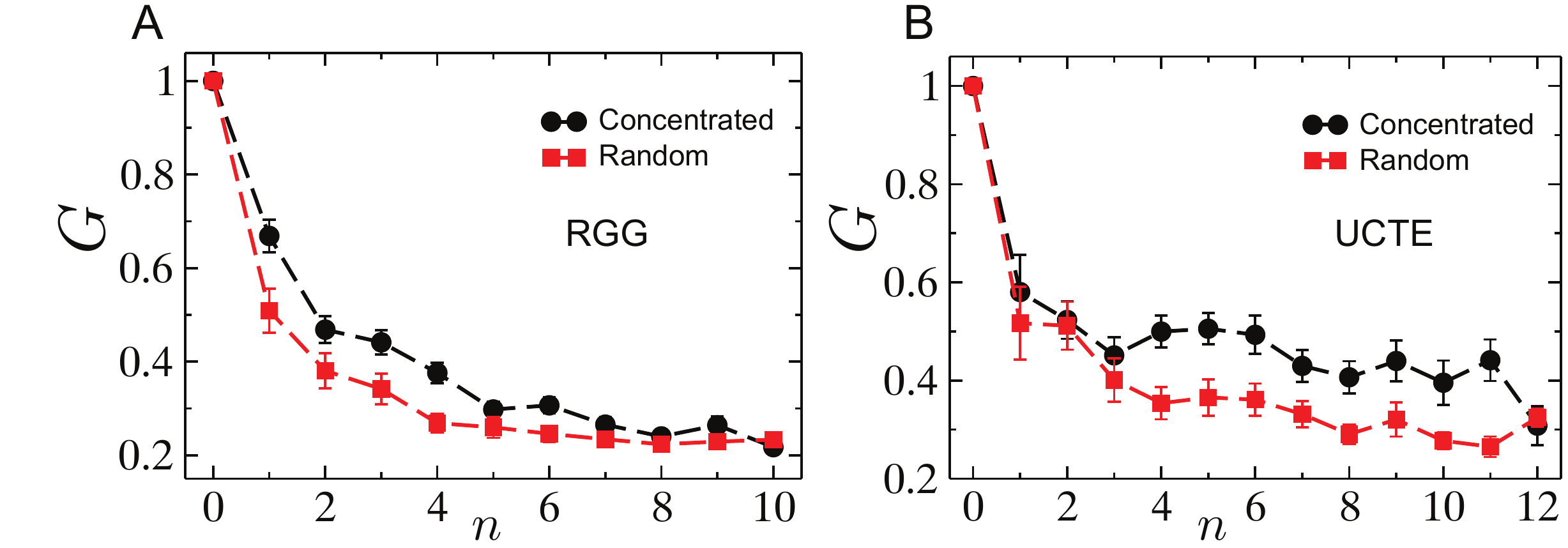}
\end{center}
\caption{{\bf Cascades triggered by concentrated versus randomly distributed removals.} (A) Fractional surviving giant component size after a cascade as a function of number of initial nodes removed in concentrated and random removals for RGGs with $N=1500$ and $\langle k \rangle = 6$.  (B) Fractional surviving giant component size after a cascade as a function of number of initial nodes removed in concentrated and random removals for the UCTE network.}
\label{fig:15}
\end{figure}

We conclude with an investigation of whether, in the case of multiple initial failures, the failures being spatially localized has any effect on the severity of the cascade. Figure~\ref{fig:15}A shows for a given value of the tolerance parameter $\alpha$, the size of the surviving giant component $G$ as a function of the number of nodes removed, for concentrated and random failures on an RGG. Random failures are only marginally more destructive than concentrated ones, which is understandable in light of how the different cascade stages resulting from just a single node's removal can cover a wide spatial spread, as seen in Fig.~\ref{fig:7}. We arrive at a similar conclusion for the case of concentrated and randomly located failures within the UCTE network from the results shown in Fig.~\ref{fig:15}B. Dynamic visualizations of the progression of spatially localized and distributed cascades on the UCTE network for the same number of initially removed nodes are provided in Supplementary Movies S3 and S4, respectively.

\section*{Discussion}

In summary, we have attempted a thorough analysis of the characteristics of cascading failures and strategies for their mitigation on spatially constrained networks, including a model of such networks viz. the random geometric graph, as well as a real-world power transmission network. The key finding worth emphasizing from these studies is the inherent lack of self-averaging for cascade processes on spatial networks. In other words, conclusions gleaned from aggregate statistics on an ensemble of such networks, yield information of little value pertaining to a single network instance. For example, in contrast to the observation for an ensemble of RGGs, for a single network instance, increasing the excess load bearing capacity does not necessarily  reduce the severity of the cascade in a monotonic fashion. Thus a straightforward measure for cascade prevention could yield counter-intuitive results. We demonstrate that increasing the effective dimensionality of the system i.e. easing the effect of the spatial constraints by introducing rewired long-range links eliminates these non-intuitive features.
A standard cascade mitigation strategy, extensively studied in the past, of preemptively removing a fraction of underperforming nodes does not effectively reduce the severity of cascades on spatially constrained networks, due to  the fairly narrow initial range of loads in spatial networks. Instead, the strategy of introducing  a fraction of altruistic nodes appears to be a more effective alternative. This holds true both for the model networks as well as for the empirical network.
Finally, we also find that cascades resulting from spatially concentrated node failures do not appear to be significantly less destructive than ones that are distributed over the network.
Thus, our results paint a complex picture for how failure cascades induced by load redistribution on spatial networks carrying distributed flow propagate through the network. In short, for spatial networks, details specific to a network instance play a very important role in determining strategies to increase the resilience of the network against cascading failures, and methods based on aggregate observations from a network ensemble will present substantial pitfalls.

{\bf Note:} Data on the UCTE network \cite {Bialek05}
that we used in this work was obtained from the website
\url{http://www.see.ed.ac.uk/~jbialek/Europe_load_flow} which is
currently non-functional. A processed version of the original data
(the UCTE network structure) can be obtained by emailing the
corresponding author (SS).



\section*{Author contributions}

Conceived and designed the experiments: AA, SS, BKS, GK.
Performed the experiments: AA, SS.
Analyzed the data: AA, SS, BKS, GK.
Wrote the manuscript: AA, SS, BKS, GK.

\section*{Supporting Information}

{\bf Supplementary Figure S1.} {\bf Cascade realizations on a single RGG of size $N=1300$ where conductances on links are inversely proportional to their lengths.} The behavior of the surviving giant component size $G$ as a function of the tolerance parameter $\alpha$ (three individual realizations are shown) is practically indistinguishable from that found in the case where conductances on all links are identical, shown in Fig.~\ref{fig:4}A in the main text. All remaining parameters (besides conductances) and simulation details are identical to that in Fig.~\ref{fig:4}A.

\vspace{0.5cm}

\noindent {\bf Supplementary Figure S2.} {\bf Effect of rewiring links in an RGG with link-length dependent conductances}. As the rewiring probability $p$ is increased, the non-monotonicities in $G$ as a function of tolerance parameter $\alpha$ gradually disappear, similarly to the case where link conductances are independent of their length (see Fig.~\ref{fig:5}). Simulations were performed with $N=1300$ and $\langle k \rangle = 5$.

\vspace{0.5cm}

\noindent {\bf Supplementary Movie S1.} {Progression of the cascade initiated by the removal of the node with the highest load on the UCTE network ($N=1254$) with tolerance parameter $\alpha = 0.4$}. Node sizes are proportional to the load on them. The single orange node at the beginning of the movie indicates the node with the largest node which is removed to trigger a cascade. The overloaded nodes in subsequent stages are shown in orange before they are removed. The total number of nodes removed in the cascade is $167$, and the number of nodes in the surviving giant component is $465$.

\vspace{0.5cm}

\noindent {\bf Supplementary Movie S2.} {\bf Progression of the cascade initiated by the removal of the node with the highest load on the UCTE network with tolerance parameter $\alpha = 0.45$.} Although the tolerance parameter is greater than in the case of Movie S1, a greater number of nodes, $299$, fail in the cascade, and the resulting giant component is also smaller, with $315$ nodes. The nodes shown in gray indicate those nodes which failed in course of the cascade occurring for $\alpha = 0.40$ (shown in Movie S1), but survived when $\alpha$ was increased to $0.45$. The survival of these nodes potentially plays a role in making the cascade more severe. All other color and node size conventions are identical to those in Movie S1.

\vspace{0.5cm}

\noindent {\bf Supplementary Movie S3.} {\bf Progression of a cascade initiated by a spatially localized removal of $9$ nodes.} Color and node size conventions are as explained in caption for Movie S1.  The tolerance parameter used here is $\alpha=0.15$. The number of nodes removed in the course of the cascade is $297$, and the number of nodes in the surviving giant component is $329$.

\vspace{0.5cm}

\noindent {\bf Supplementary Movie S4.} {\bf Progression of a cascade initiated by distributed (random) removal of $9$ nodes.} Color and node size conventions are as explained in caption for Movie S1. The tolerance parameter used here is $\alpha=0.15$. The number of nodes removed in the course of the cascade is $297$ (same as for Movie S3), and the number of nodes in the surviving giant component is $374$.

\renewcommand{\figurename}{Supporting Figure}
\setcounter{figure}{0}

\newpage 

\begin{center}{\bf \Large SUPPLEMENTARY FIGURES}\end{center}


\begin{figure}[!ht]
\begin{center}
\includegraphics[width=4.5in]{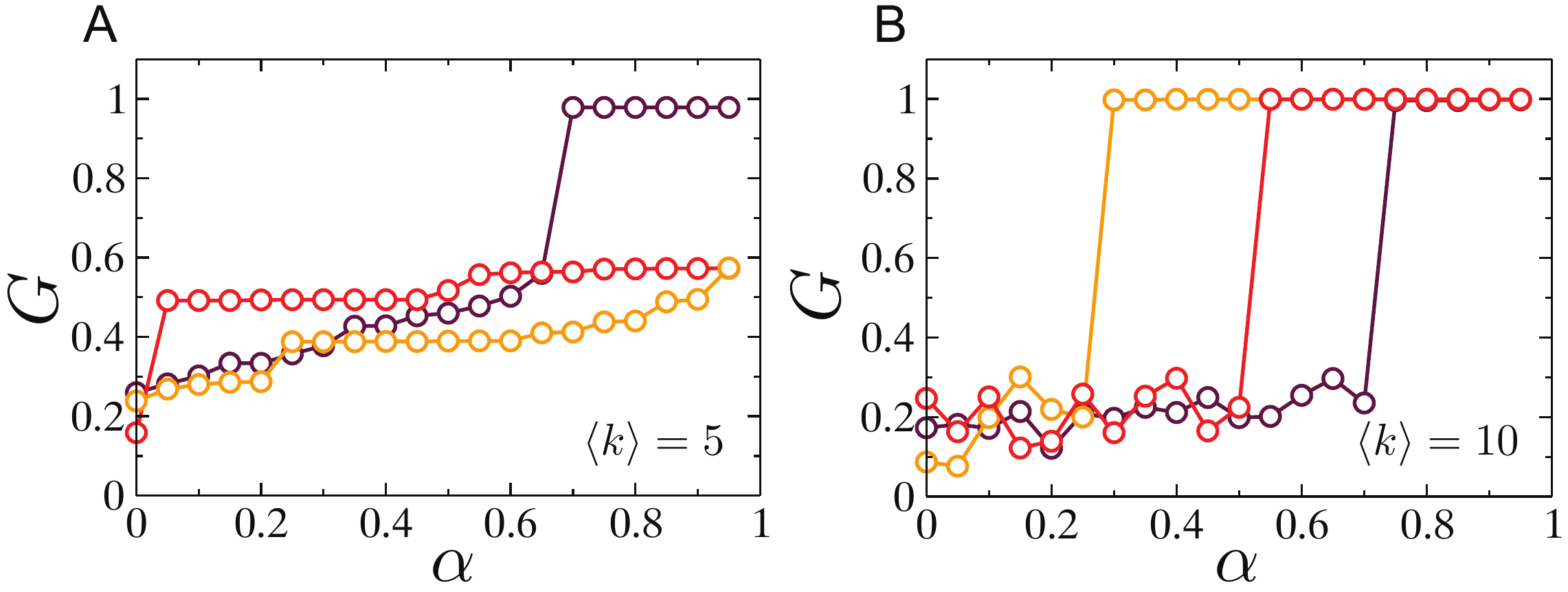}
\end{center}
\caption{ {\bf Cascade realizations on a single RGG of size $N=1300$ where conductances on links are inversely proportional to their lengths.} The behavior of the surviving giant component size $G$ as a function of the tolerance parameter $\alpha$ (three individual realizations are shown) is practically indistinguishable from that found in the case where conductances on all links are identical, shown in Fig.~\ref{fig:4}A in the main text. All remaining parameters (besides conductances) and simulation details are identical to that in Fig.~\ref{fig:4}A.
}
\label{fig:S1}
\end{figure}

\begin{figure}[!ht]
\begin{center}
\includegraphics[width=4.5in]{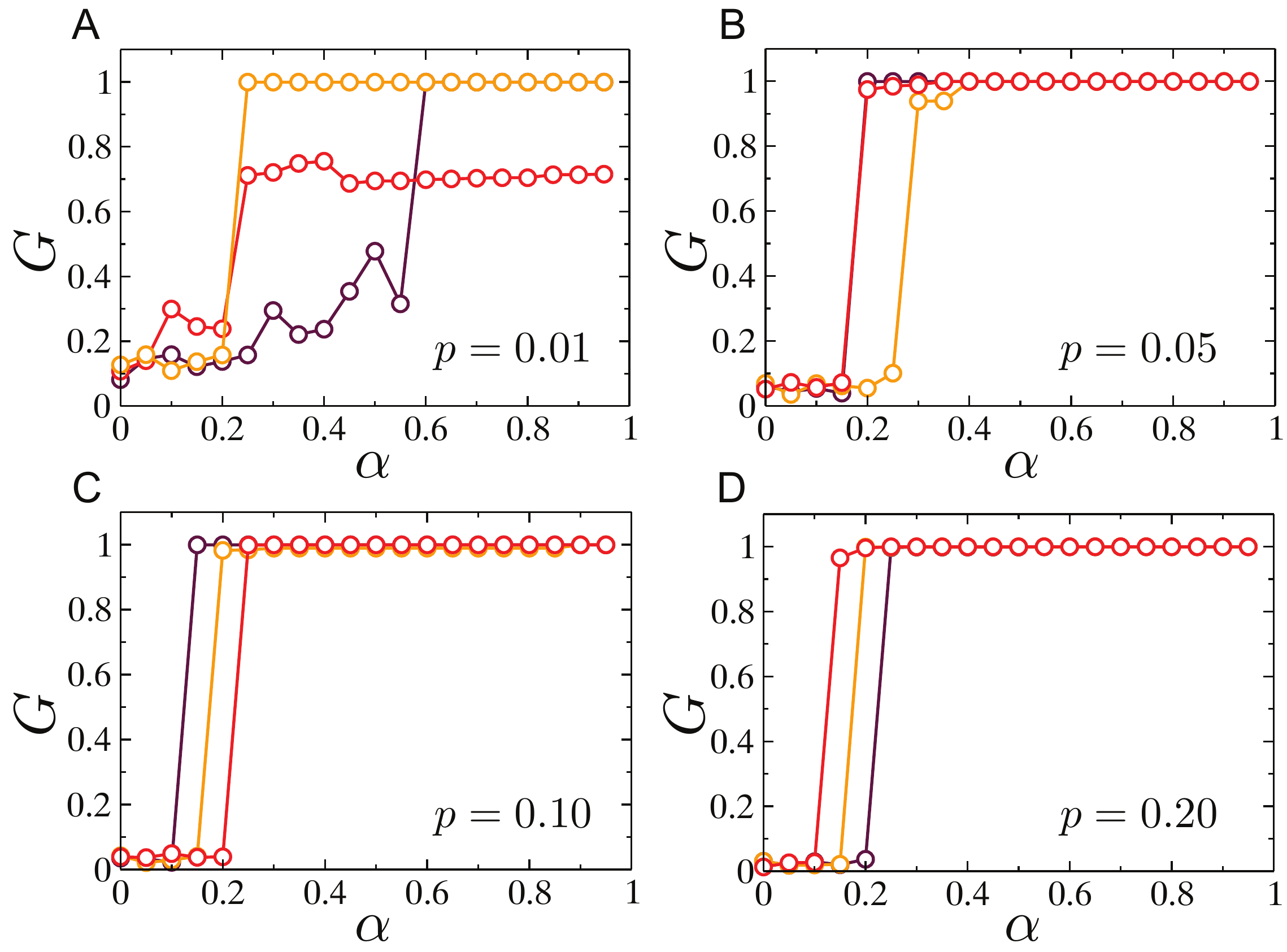}
\end{center}
\caption{ {\bf Effect of rewiring links in an RGG with link-length dependent conductances}. As the rewiring probability $p$ is increased, the non-monotonicities in $G$ as a function of tolerance parameter $\alpha$ gradually disappear, similarly to the case where link conductances are independent of their length (see Fig.~\ref{fig:5}). Simulations were performed with $N=1300$ and $\langle k \rangle = 5$.
}
\label{fig:S2}
\end{figure}

\end{document}